\documentclass[structabstract]{aa}  
\usepackage{natbib}
\usepackage{graphicx}
\usepackage{txfonts}
\begin{document}
   \title{Star Formation in Extreme Environments: The Effects of Cosmic Rays and Mechanical Heating.}


   \author{R. Meijerink\inst{1} \and M. Spaans\inst{2} \and A.F. Loenen\inst{1} \and P.P. van der Werf\inst{1}
   }

   \institute{Leiden Observatory, Leiden University, P.O. Box 9513,
     NL-2300 RA Leiden, Netherlands\\ 
     \email{meijerink@strw.leidenuniv.nl} \and
     Kapteyn Astronomical Institute, PO Box 800, 9700 AV Groningen, The Netherlands\\
     \email{spaans@astro.rug.nl}
    }
   \date{Received ??; accepted ??}

  \abstract{Molecular data of extreme environments, such as Arp 220,
    but also NGC~253, show evidence for extremely high cosmic ray (CR)
    rates ($10^3-10^4\times$ Milky Way) and mechanical heating from
    supernova driven turbulence.}{The consequences of high CR rates
    and mechanical heating on the chemistry in clouds are
    explored.}{PDR model predictions are made for low, $n=10^3$, and
    high, $n=10^{5.5}$~cm$^{-3}$, density clouds using well-tested
    chemistry and radiation transfer codes. Column densities of
    relevant species are discussed, and special attention is given to
    water related species. Fluxes are shown for fine-structure lines
    of O, C$^+$, C, and N$^+$, and molecular lines of CO, HCN, HNC,
    and HCO$^+$. A comparison is made to an X-ray dominated region
    model.}{Fine-structure lines of [CII], [CI], and [OI] are
    remarkably similar for different mechanical heating and CR rates,
    when already exposed to large amounts of UV. HCN and H$_2$O
    abundances are boosted for very high mechanical heating rates,
    while ionized species are relatively unaffected. OH$^+$ and
    H$_2$O$^+$ are enhanced for very high CR rates $\zeta \ge 5\cdot
    10^{-14}$~s$^{-1}$. A combination of OH$^+$, OH, H$_2$O$^+$,
    H$_2$O, and H$_3$O$^+$ trace the CR rates, and are able to
    distinguish between enhanced cosmic rays and X-rays.}{}


   \maketitle

%

\section{Introduction}

Observations of molecular tracers have suggested that both star
formation and an active galactic nucleus (AGN, an accreting
supermassive black hole) can drive the physics and chemistry of the
central ~kpc of active galaxies \citep{Sanders1996, Gao1997, Gao1999,
  Baan2008, VdWerf2010}. Specifically, Ultra-Luminous Infrared
Galaxies (ULIRGs), with infrared ($8-1000$~$\mu$m) luminosities
$L_{\rm IR}\ge 10^{12}$ L$_\odot$, appear to contain large reservoirs
of dense, $n>10^4$ cm$^{-3}$, gas. This gas provides the fuel for
(intense) star formation and black hole accretion.

The irradiation by UV photons \citep[6-13.6 eV; creating so-called
  photon-dominated regions, or PDRs, e.g.,][]{Hollenbach1999} from a
starburst or by X-rays \citep[1-100 keV, creating so-called X-ray
  dominated regions, or XDRs, e.g.,][]{Lepp1996, Maloney1996} from an
AGN is expected to produce different signatures in the molecular
chemistry \citep{Meijerink2005, Meijerink2006,
  Meijerink2007}. Recently, mechanical heating due to supernova and
stellar outflow driven turbulence has been explored as well
\citep{Loenen2008}, which shows that warm molecular gas may persist
beyond a few magnitudes of extinction in molecular
clouds. Furthermore, the radiation feedback from ultraviolet photons,
resulting in warm dust of more than 50 K, affect the thermodynamics of
the molecular gas, possibly resulting in a top-heavy initial mass
function \citep[IMF;][]{Klessen2007, Hocuk2010}.

The analysis of the physical state of this irradiated gas allows one
to constrain the evolutionary state of, ambient star formation rate
in, and the importance of feedback for ULIRGs \citep{Baan2010}. In
this work, we focus on feedback that results from massive stars. In
particular, we consider UV emission from hot stars, the (elevated)
production of cosmic rays from the resulting supernova remnants, and
the heating that results from turbulent motions that are driven by
travelling supernova blast waves.

A very nice case in point for the feedback effects mentioned above is
Arp 220, the proto-typical ULIRG at 77 Mpc, which boasts a infrared
luminosity of $L_{\rm IR}=1.3\cdot 10^{12}$ L$_\odot$
\citep{Soifer1987}. Arp 220 is a merger and contains two nuclei
separated by about 370 pc in the east-west direction
\citep[e.g.,][]{Rodriguez2005}. The nuclear star formation rate of Arp
220 is 50-100 M$_\odot$ yr$^{-1}$ \citep{Smith1998}.

Observations of Arp 220 at a resolution of $0.3\arcsec$ (100 pc) by
\citet{Sakamoto2009} indicate P Cygni type profiles in HCO$^+$ 4-3,
3-2 and CO 3-2. These authors interpret their results as a $\sim 100$
km s$^{-1}$ outflow that is driven by supernova explosions (or
radiation pressure from the starburst). Interestingly, as many as 49
supernova remnants (SNRs) have been detected with VLBI (1 pc
resolution) at 18 cm by \citep{Lonsdale2006}, confirming the
formation and demise of many massive stars. The latter authors derive
a supernova rate of 3 yr$^{-1}$ for the western nucleus and 1
yr$^{-1}$ for the eastern nucleus, more than two orders of magnitude
larger than the entire Milky Way.

Limits on the cosmic ray (CR) rate in our Milky Way are discussed by,
e.g., \citet{VdTak2000}. They found a value of $\zeta=2.6\pm0.8\cdot
10^{-17}$~s$^{-1}$ with upper limits that are 3 to 5 times higher. In
this paper we adopt $\zeta=5\cdot 10^{-17}$~s$^{-1}$ as the canonical
Milky Way value. Highly elevated cosmic ray rates are seen in the
recent Gamma ray observations of the starburst galaxy NGC~253
\citep{Acero2009} with the High Energy Stereoscopic System (H.E.S.S.)
array. The observed gamma ray flux indicates a cosmic ray density that
is three orders of magnitude larger than our Milky Way.

\citet{Papadopoulos2010} suggests that ULIRGs generally have cosmic
ray densities $U_{\rm CR}\sim 10^3 - 10^4 U_{\rm cr, MW}$, and
discusses the impact on the ionization fraction and thermal state of
the gas, and the consequences for the initial conditions for
star-formation. In this paper, we first discuss the detailed effects
of enhanced cosmic ray rates on both the thermal balance and chemistry
(Section \ref{Chemistry}). Then, we will highlight the global
consequences on the chemistry and suggest potential diagnostic species
(Section \ref{Col_dens}), which is followed by the observational
consequences (Section \ref{Obs_diag}). In Section \ref{key_diagn}, we
will conclude with the key diagnostics needed to derive quantitive and
qualitative properties of interstellar medium (ISM) exposed to large
amounts of cosmic rays, and show how it differs from gas exposed
X-rays.

\section{PDR models}\label{Chemistry}

We have constructed a set of PDR models (see Table \ref{model_params})
from the code as described by \citet{Meijerink2005} and
\citet{Meijerink2007}. We consider a low ($n=10^3$~cm$^{-3}$), and a
high density ($n=10^{5.5}$~cm$^{-3}$) phase. The low density phase of
the ISM is generally not directly connected to
starformation. Therefore, the adopted incident UV field in the low
density models ($G_0=10^3$) is chosen a factor 100 lower than the high
density phase ($G_0=10^5$) in order to reflect this and essentially
corresponding to Model 1 and 4 in \citet{Meijerink2005}. We varied the
incident cosmic ray rates between $\zeta=5\cdot 10^{-17}$~s$^{-1}$ and
$\zeta=5\cdot 10^{-13}$~s$^{-1}$. The highest cosmic ray rate
considered here is two orders of magnitude higher than in
\citet{Meijerink2006}. We adopt a Solar metallicity $Z=Z_\odot$ and an
abundance ratio ${\rm [C]/[O]}=0.4$, which is based on the average
abundances of \citet{Asplund2005} and \citet{Jenkins2004}, and
summarized in Table 2 of Meijerink \& Spaans \citeyear{Meijerink2005}.

We also consider the effect of mechanical heating in the high density
case, specifically $\Gamma_{\rm mech}=3.0\cdot 10^{-18}$ and $2.0\cdot
10^{-17}$~erg~cm$^{-3}$~s$^{-1}$, which correspond to star formation
rates of about $SFR =140$ and $950$~M$_\odot$~yr$^{-1}$, respectively,
for a Salpeter IMF \citep[see][for details]{Loenen2008}. Mechanical
heating is only considered in the high density case, $n>10^4-10^5$
cm$^{-3}$, since we expect this dense phase to be directly responsible
for star formation and thus be affected directly by the supernovae
from massive stars. We assume that the low density phase is only
affected by cosmic rays. These are able to travel through large
columns ($N_{\rm H}\sim 10^{24}$~cm$^{-2}$), since they have small
absorption cross sections, and are likely to impact the ISM on larger
scales. 

The effects on the thermal balance and chemistry are discussed for
Model sets 1a ($G_0=10^5$; $n=10^{5.5}$~cm$^{-3}$), 1b and 1c (same as
Model set 1a, with $\Gamma_{\rm mech}=3.0\cdot 10^{-18}$ and $2.0\cdot
10^{-17}$~erg~cm$^{-3}$~s$^{-1}$, respectively) and 2 ($G_0=10^3$;
$n=10^3$~cm$^{-3}$) in order to highlight the differences between the
high and low density phases of the ISM. 

\begin{table}
\caption{Model parameters}   
\label{model_params}    
\centering         
\begin{tabular}{l c c c}    
\hline\hline                
Metallicity [Z$_\odot$]     & \multicolumn{3}{l}{1.0} \\
$[$C$]$:$[$O$]$ ratio                  & \multicolumn{3}{l}{0.4} \\
CR rates [s$^{-1}$] & \multicolumn{3}{l}{$5\cdot 10^{-17}$, $5\cdot 10^{-16}$, $5\cdot 10^{-15}$, 
$5\cdot 10^{-14}$, $5\cdot 10^{-13}$} \\
\hline
Model set  & Density    & Radiation field & $\Gamma_{\rm mech}$ \\  
           & [cm$^{-3}$] & [$G_0$]         & [erg cm$^{-3}$~s$^{-1}$]  \\
\hline                     
1a         & 10$^{5.5}$  & 10$^5$          & 0.0    \\
1b         & 10$^{5.5}$  & 10$^5$          & $3.0\cdot 10^{-18}$ \\
1c         & 10$^{5.5}$  & 10$^5$          & $2.0\cdot 10^{-17}$ \\
2          & 10$^3$     & 10$^3$          & 0.0 \\ 
\hline
\end{tabular}
\end{table}

\subsection{Thermal balance}

The surface temperature for the high density model without boosting
the CR rate (i.e. $\zeta=5\cdot 10^{-17}$~s) is higher than for the
low density model. This may seem a little counterintuitive, since
$G_0/n$=1 (0.3) for the low (high) density model, and as a result
there is more energy available per particle in the low density
model. Besides that, cooling is more efficient at high
densities. However, at high radiation fields ($G_0>10^4$, and
$n>10^4-10^5$~cm$^{-3}$), the gas cooling thermalizes
($\Lambda~\propto n$ instead of $n^2$), while photo-electric heating
becomes more effective because of the higher electron density, and
more effective than $\Gamma \sim n$ \citep[e.g.,][]{Kaufman1999,
  Meijerink2007}.

\begin{figure*}
  \centering
  \includegraphics[width=18cm,clip]{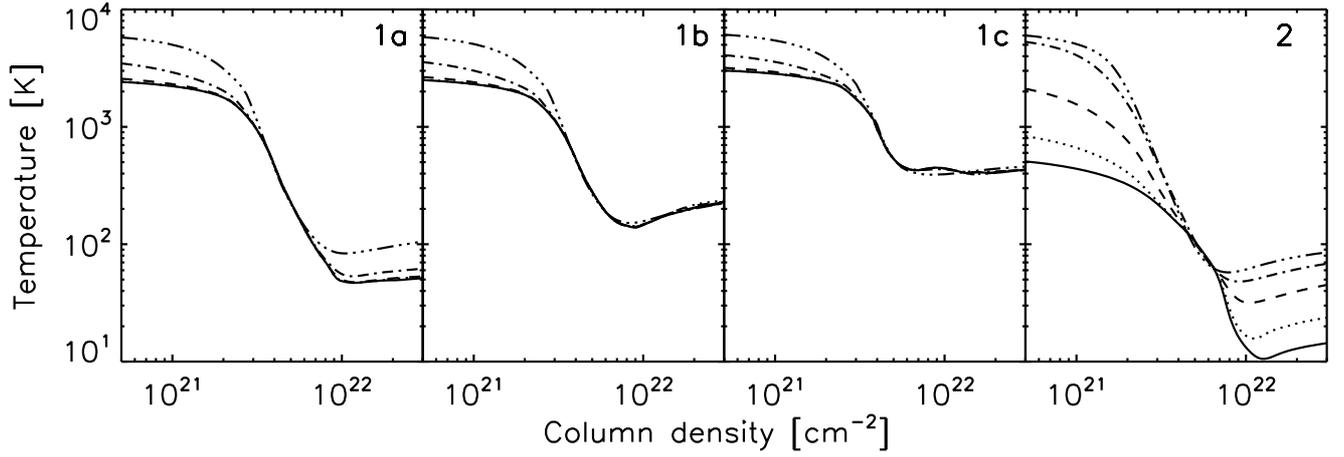}
  \caption{Temperatures for all models sets (see Table
    \ref{model_params}) for cosmic ray rates $\zeta=5\cdot 10^{-17}$
    (solid line), $5\cdot 10^{-16}$ (dotted), $5\cdot 10^{-15}$
    (dashed), $5\cdot 10^{-14}$ (dot-dashed), and $5\cdot
    10^{-13}$~s$^{-1}$ (dotted-dashed).}
  \label{Model_temperatures}
\end{figure*}

The sensitivity of the thermal balance to cosmic rays is different for
the high and low density phase (see Fig. \ref{Model_temperatures}). In
Model set 1a, only very high CR rates ($\zeta \geq 5\cdot
10^{-14}$~s$^{-1}$) are able to increase the temperature significantly
($\sim 25$ to 200 percent) in both the radical region (low $A_V$) as
well as the shielded molecular region. In Model sets 1b and 1c,
mechanical heating in the shielded region of the cloud dominates over
cosmic ray heating. Here, the temperatures overlap for all cosmic ray
rates, reaching values $T\sim 100-200$~K for Model set 1b and $T\sim
400$~K for Model set 1c. So even for moderate mechanical heating rates
(in this case corresponding to a SFR=140 $M_\odot/$yr), the CR rate
has no additional effect on the thermal balance in the shielded
regions of clouds. In Model set 2, temperatures are increased for all
CR rates at all column densities, up to an order of magnitude with the
highest adopted CR rate. The effect on the thermal balance is larger,
because cooling is simply less efficient at lower densities ($\Lambda
\propto n^2$), compared to the heating rate that is proportional to
the density, ($\Gamma \propto n$). This makes the low density models
more sensitive to a larger heating rate. This is enhanced by the fact
that the fractional electron abundance is increased more easily as
well at lower densities ($k_{rec}\propto n^2$), resulting in a higher
heating efficiency \citep[cf.,][]{Bakes1994}, and higher heating
rates.

\subsection{Chemical balance}

\begin{figure*}
  \centering
  \includegraphics[width=18cm,clip]{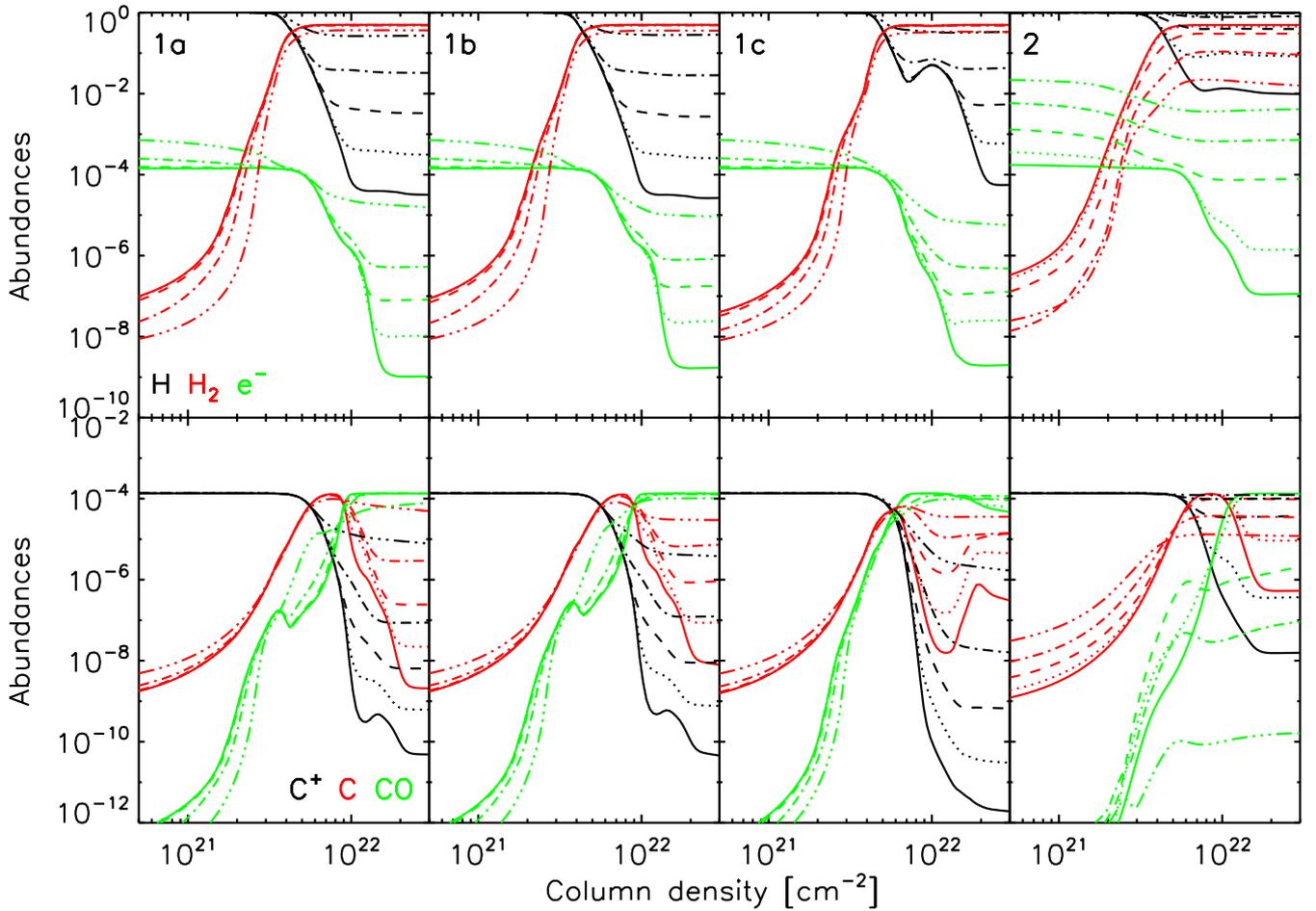}
  \caption{Abundances of H, H$_2$, e$^-$, C$^+$, C and CO for all
    models sets (see Table \ref{model_params}). Line styles
    corresponding to the different CR rates are the same as in
    Fig. \ref{Model_temperatures}.}
  \label{Model_abun_H_H2_e_Cp_C_CO}
\end{figure*}

\textbf{\textit{Electron abundances
    (Fig. \ref{Model_abun_H_H2_e_Cp_C_CO}):}} The effect of cosmic
rays is less apparent in the unshielded region ($N_{\rm H} \lesssim
3\cdot 10^{21}$~cm$^{-2}$) than in the molecular region ($N_{\rm
  H}\gtrsim 10^{22}$~cm$^{-2}$) of the cloud, since charged species in
the unattenuated part of cloud are also produced through
photoionization by the stellar radiation field. Here, the fractional
abundance (which is defined as $x_{\rm i} = n_{\rm i} / n({\rm
  H+2H_2}$) of electrons varies between $x_{\rm e}\sim 10^{-4} -
10^{-3}$ ($10^{-4}-10^{-2}$) in the high (low) density models, while
the CR rates vary over 4 orders of magnitude. In the molecular region,
however, an order of magnitude higher CR rate gives a similar rise in
the electron fractional abundance, giving ranges $x_{\rm e}\sim
10^{-9}-10^{-5}$ and $10^{-7}-10^{-2}$ for the high and low density
models, respectively. Adding mechanical heating (1b and 1c) does not
change the electron abundance much. The difference in fractional
electron abundance between the highest and lowest CR rate is somewhat
larger than for the models without mechanical heating.

\textbf{\textit{H and H$_2$ abundances (Fig. \ref{Model_abun_H_H2_e_Cp_C_CO}):}} Atomic hydrogen is by far
the dominant species in the radical region ($N_{\rm H} \lesssim
2-3\cdot 10^{22}$~cm$^{-2}$), but when the attenuating column becomes
larger, molecular hydrogen becomes more and more abundant. For the
canonical CR rate ($\zeta=5\cdot 10^{-17}$~s$^{-1}$), the atomic
hydrogen abundance drops below $x_{\rm H}\sim 10^{-4}$ ($10^{-2}$) for
the high (low) density models. The H to H$_2$ transition becomes less
and less complete for higher CR rates, which is due to CR reactions
such as $\rm H_2 + CR \rightarrow H + H^+ + e^-$ and $\rm H_2 + CR
\rightarrow H_2^+ + e^-$. Atomic and molecular hydrogen are of similar
abundance for the highest adopted CR rate in the high density case,
while atomic hydrogen is dominant for CR rates $\zeta > 5\cdot
10^{-15}$~s$^{-1}$ in the low density case.

\textbf{\textit{C$^+$, C and CO abundances
    (Fig. \ref{Model_abun_H_H2_e_Cp_C_CO}):}} Similar trends are seen
for the C$^+$/C/CO transition. For the lowest CR rate ($\zeta=5\cdot
10^{-17}$~s$^{-1}$), there are three well defined zones where one
species is most abundant. Increasing CR rates result in a larger
neutral carbon abundance in the unattenuated part of the cloud by a
factor of $\sim3$ ($\sim30$) in the high (low) density case. However,
the abundances are still low compared to C$^+$. Again most effect is
seen at large column density ($N_{\rm H}>10^{22}$~cm$^{-2}$). The
C$^+$ and C abundances increase very fast with the CR rate. In the
high density case, the C$^+$ and C abundances become $10^{-6}-10^{-5}$
and $3-6\cdot 10^{-5}$, respectively, depending on the amount of
additional mechanical heating. There is no full transition to CO for
$\zeta \geq 10^{-15}$~s$^{-1}$ in the low density case. Neutral carbon
is the dominant species at high column density $\zeta = 5\cdot
10^{-15}$~s$^{-1}$, and C$^+$ has become the dominant species at
$\zeta \geq 5\cdot 10^{-14}$~s$^{-1}$, and at the same time there is
almost no CO left ($x_{\rm CO} < 10^{-6}$).

\begin{figure*}
  \centering
  \includegraphics[width=18cm,clip]{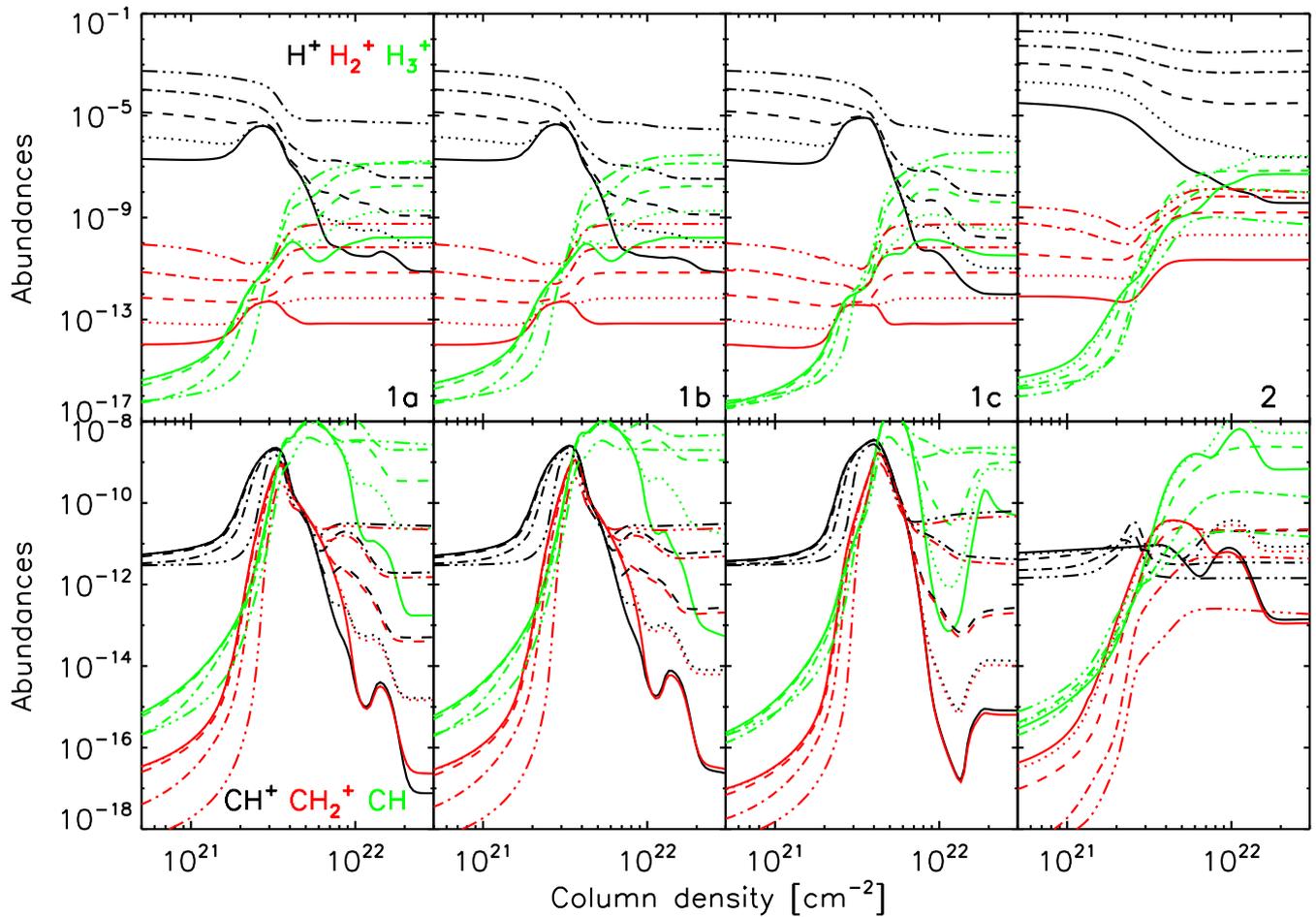}
  \caption{Abundances of H$^+$, H$_2^+$, H$_3^+$, CH, CH$^+$, and
    CH$_2^+$ for all model sets (see Table \ref{model_params}). Line
    styles corresponding to the different CR rates are the same as in
    Fig. \ref{Model_temperatures}.}
  \label{Model_abun_Hp_H2p_H3p_CH_CHp_CH2p}
\end{figure*}

\textbf{\textit{H$^+$, H$_2^+$, and H$_3^+$ abundances
    (Fig. \ref{Model_abun_Hp_H2p_H3p_CH_CHp_CH2p}):}} The main drivers
of chemistry in PDRs are ionic species, starting with species H$^+$,
H$_2^+$, and H$_3^+$. H$^+$ and H$_2^+$ are mainly (directly or
indirectly) produced by cosmic ray ionization ($\rm H + CR \rightarrow
H^+ + e^-$; $\rm H_2 + CR \rightarrow H_2^{+} + e^-$ or $\rm H^+ + H +
e^-$ or $\rm H + H$). UV photons of energy larger than 13.6 eV are all
readily absorbed in a very small column, forming an HII region in
front of the PDR. H$_2$ can only be photodissociated to two hydrogen
atoms (with an UV excited H$_2^*$ as intermediate), no ionic species
are produced by direct photoionization of H$_2$. Therefore, large
effects on abundances of ionic species are seen by varying the cosmic
ray rate.

In the unshielded region, the abundance increase of H$^+$ and H$_2^+$
scale with the ionization rate. H$^+$ is directly produced by cosmic
ray ionization. The H$_2^+$ abundance is balanced by the charge
exchange reactions $\rm H^+ + H_2 \leftrightarrow H_2^+ + H$. The
reaction to the right has an energy barrier of $T\sim 20000$~K, and is
only efficient at temperatures of a few thousand Kelvin. Therefore,
the H$_2^+$ abundance is still able to increase with CR rate: the
smaller H$_2$ abundance at higher CR rates is counteracted by a
temperature increase (see Fig. \ref{Model_temperatures}). At slightly
larger column densities ($N_{\rm H} \sim 3\cdot 10^{21}$~cm$^{-2}$),
the temperature drops and only $\rm H_2^+ + H \rightarrow H^+ + H_2$
is efficient, while simultaneously the H to H$_2$ transition occurs. This
together results in an abundance maximum of H$^+$, and also explains
why the H$_2^+$ abundance is not increasing as fast as the H$_2$
abundance.

After the H to H$_2$ transition ($N_{\rm H} \gtrsim 5 \cdot
10^{21}$~cm$^{-2}$), the overall ionization degree drops, which
coincides with a decrease in temperature. This is mostly explained by
recombination reactions (e.g., $\rm H^+ + e^- \rightarrow H +
Photon$), that are more efficient at lower temperatures. In this
region, the increase of H$^+$ with cosmic ray rate is faster than
linear. The H to H$_2$ transition is less complete for higher CR rates
(Fig. \ref{Model_abun_H_H2_e_Cp_C_CO}), giving a larger reservoir of
atomic hydrogen, that can potentially be ionized.

H$_3^+$ is formed and destroyed by $\rm H_2^+ + H_2 \rightarrow H_3^+
+ e^-$ and $\rm H_3^+ + e^- + \rightarrow 3H$ and $\rm H_2 + H$,
respectively. At low column densities, the H$_3^+$ abundance is
dropping with CR rates due to enhanced electron and reduced H$_2$
abundances. At higher column densities, H$_3^+$ increases for large CR
rates, but not one to one with the CR rate, due to more efficient
recombination and lower H$_2$ abundances (especially in the low
density models).

\textbf{\textit{CH$^+$, CH$_2^+$, and CH abundances
    (Fig. \ref{Model_abun_Hp_H2p_H3p_CH_CHp_CH2p}):}} In the
unshielded region of the cloud, the formation of CH$^+$ occurs mainly
through $\rm C^+ + H_2^* \rightarrow CH^+ + H$, where $\rm H_2^*$ is
vibrationally excited molecular hydrogen. Therefore, the CH$^+$
abundances do not change much with cosmic ray rate in the unshielded
region of the cloud. However, at larger column densities ($N_{\rm H}
\sim 10^{22}$~cm$^{-2}$), the main route is through $\rm H_3^+ +
C$. At high density, both the H$_3^+$ and C abundance increase orders
of magnitude for the larger CR rates, and as a result the CH$^+$
abundance as well, despite the increased electron abundance. In the
low density case, the CH$^+$ abundances are very similar for all CR
rates because there is no boost now through C + H$_3^+$. CH$_2^+$
formation occurs mainly through $\rm CH^+ + H_2 \rightarrow CH_2^+ +
H$ and follows the abundance of CH$^+$ closely, except where molecular
hydrogen is not the dominant species, i.e., the radical region in the
high density case and for very high CR rates in the low density case.

CH can be formed by the neutral-neutral reactions $\rm C + H_2
\rightarrow CH + H$ and $\rm C + OH \rightarrow CH + O$, which have
energy barriers of $T \sim 11700$ and 14800~K, respectively, and the
recombination reaction $\rm CH_2^+ + e^- \rightarrow CH +
H$. Destruction happens through UV photodissociation and
photoionization at small column densities, and by fast charge exchange
reactions (e.g., $\rm H^+ + CH \rightarrow CH^+ + H$) at all column
densities. At small column densities ($N_H \lesssim
10^{21}$~cm$^{-2}$), the CH abundances are fairly similar, and UV
photoionization dominates the destruction, and neutral-neutral
reactions formation. The small differences can be explained by small
deviations in H$_2$ and C abundances. At column densities $N_{\rm H}
\gtrsim 5\cdot 10^{21}$~cm$^{-2}$, the formation is dominated by
recombination of CH$_2^+$, and the abundance is mainly determined by
the CR rate.

\begin{figure*}
  \centering
  \includegraphics[width=18cm,clip]{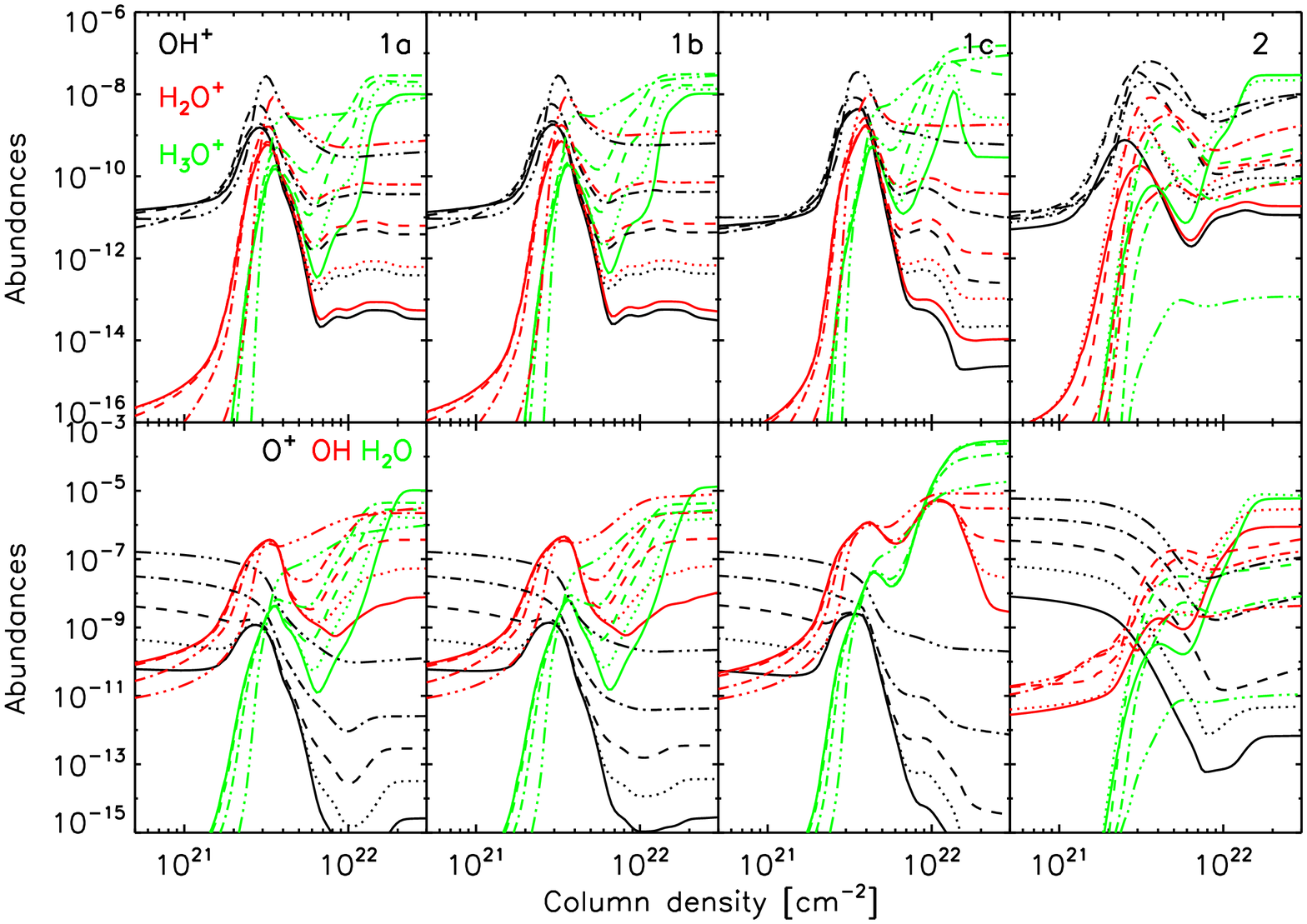}
  \caption{Abundances of OH$^+$, H$_2$O$^+$, H$_3$O$^+$, O$^+$, OH,
    and H$_2$O for all models sets (see Table
    \ref{model_params}). Line styles corresponding to the different CR
    rates are the same as in Fig. \ref{Model_temperatures}.}
  \label{Model_abun_OHp_H2Op_H3Op_Op_OH_H2O}
\end{figure*}

\textbf{\textit{Water chemistry
    (Fig. \ref{Model_abun_OHp_H2Op_H3Op_Op_OH_H2O}):}} H$_2$O and
related species, OH$^+$, OH, H$_2$O$^+$, and H$_3$O$^+$ are
significantly affected by cosmic rays. First we consider H$_2$O. It
can be formed through the neutral-neutral reaction $\rm H_2 + OH
\rightarrow H_2O + H$, for which temperatures $T > 200-300$~K are
required. Cosmic rays are not able to increase the temperature to
these temperatures, not even the highest cosmic ray rates. Another
important route, especially when gas is significantly ionized by
enhanced cosmic ray rates, is, e.g., $\rm H_2^+ + O \rightarrow OH^+ +
H$, then $\rm OH^+ + H_2 \rightarrow H_2O^+ + H$, $\rm H_2O^+ + H_2
\rightarrow H_3O^+ + H$, followed by recombination, $\rm H_3O^+ + e^-
\rightarrow H_2O + H$. Destruction is possible through UV
photodissociation, charge exchange, proton transfer, ion-molecule
reactions and cosmic rays. Important routes are $\rm He^+ + H_2O
\rightarrow OH + He + H^+$ and $\rm H^+ + H_2O \rightarrow H_2O^+ +
H$, and varying the cosmic ray rates affect these
abundances. Therefore, at intermediate column densities $N_H \sim
3\cdot 10^{21}$ to $10^{22}$~cm$^{-2}$ in model sets 1a and 1b, the
water abundances increase for higher CR rates. In these models, the
temperatures are not high enough to form H$_2$O through
neutral-neutral reactions. When the temperature is high enough to form
water (model set 1c) through the neutral-neutral reactions, a higher
ionization degree is a drawback (H$_2$O is destroyed by, e.g.,
He$^+$), and the water abundance decreases for very high CR rates
($\zeta \gtrsim 5\cdot 10^{-14}$~s$^{-1}$). In model set 2, there is
not enough H$_2$ at the higher cosmic ray rates to form water through
ion-molecule reactions.

OH is formed through $\rm H_2 + O \rightarrow OH + H$ or destruction
of water (see previous paragraph). In the radical region where
temperatures are signficant, formation through neutral reactions
dominates. In the high density models they are closely connected to
the H$_2$ abundance, and here the OH abundance is slightly lower for
the high CR rates. In the low density case, the OH abundance increases
with CR rate in the radical region, since the temperature increases
about an order of magnitude. When temperatures are lower, i.e.,
$N_H\gtrsim 5\cdot 10^{21}$~cm$^{-2}$, the contribution of water
destruction becomes more important. OH abundances are significantly
larger in model set 1a and 1b, when CR rates are higher. In model set
1c, formation through neutral-neutral reactions becomes more
important, and the effect of an enhanced CR rate less evident.

OH$^+$ is formed by $\rm H^+ + OH \rightarrow OH^+ +H$ and $\rm H^+ +
O \rightarrow O^+ + H$ followed by $\rm O^+ + H_2 \rightarrow OH^+ +
H$ or by photoionization of OH. At small column densities, where
OH$^+$ abundances are very similar for different CR rates, the
abundance is determined by the UV ionization of OH. At large column
densities ($N_{\rm H} \gtrsim 5\cdot 10^{21}$~cm$^{-2}$ ), the UV is
mostly gone, and here the H$^+$ driven formation is more important,
which is enhanced by cosmic rays. $\rm H_2O^+$ closely follows the
OH$^+$ abundance at these high column densities. It is mainly formed
through $\rm H_2 + OH^+ \rightarrow H_2O^+ + H$, and also very much
enhanced at higher CR rates.

\begin{figure*}
  \centering
  \includegraphics[width=18cm,clip]{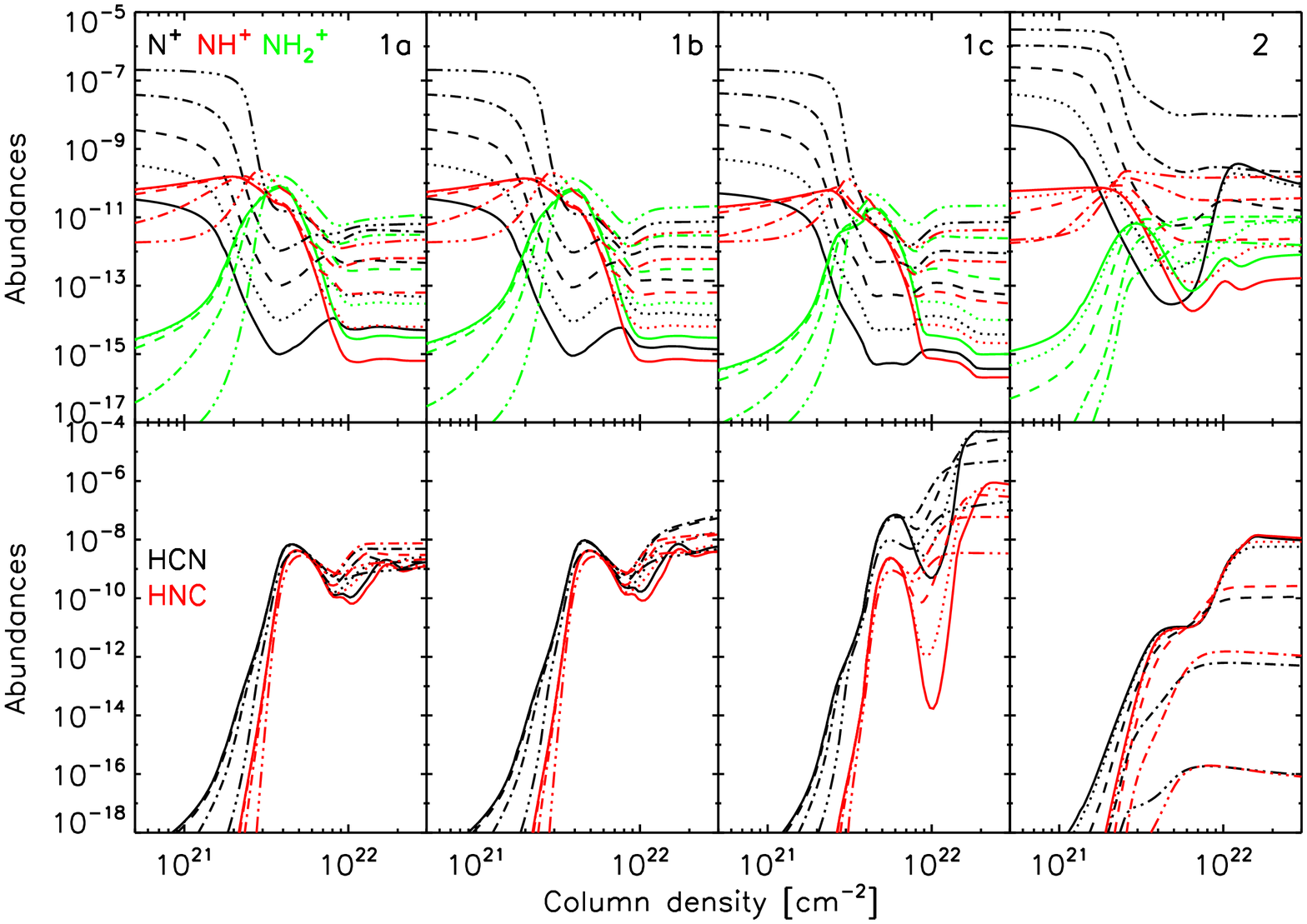}
  \caption{Abundances of N$^+$, NH$^+$, NH$_2^+$, HCN, and HNC for all
    models sets (see Table \ref{model_params}). Line styles
    corresponding to the different CR rates are the same as in
    Fig. \ref{Model_temperatures}.}
  \label{Model_abun_Np_NHp_NH2p_HCN_HNC}
\end{figure*}

\begin{figure*}
  \centering
  \includegraphics[width=18cm,clip]{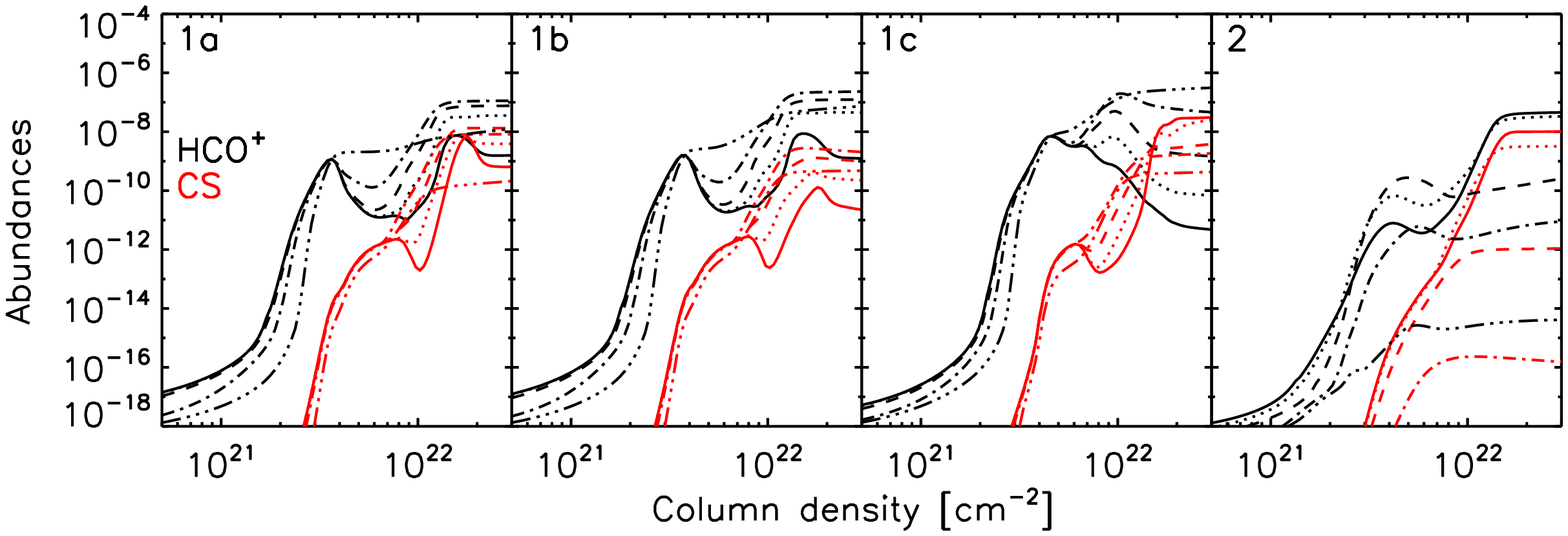}
  \caption{Abundances of HCO$^+$ and CS for all model sets (see Table
    \ref{model_params}). Line styles corresponding to the different CR
    rates are the same as in Fig. \ref{Model_temperatures}.}
  \label{Model_abun_HCOp_CS}
\end{figure*}

\textbf{\textit{Nitrogen chemistry
    (Fig. \ref{Model_abun_Np_NHp_NH2p_HCN_HNC}):}} N$^+$ has an
ionization potential of 14.53 eV, which is too high for the available
UV photons. This species shows orders of magnitude variations for all
model sets. The species is strongly affected by the increased cosmic
ray rate. NH$^+$ and NH$_2^+$ are closely connected to the cosmic ray
rate at high column densities, because the formation is driven by $\rm
N + H_2^+$ and $\rm N^+ + H_2$.

HCN and HNC do not show a strong response to enhanced CR rates in the
high density case without mechanical heating, and the abundances
remain similar for all CR rates. Adding mechanical heating (1b),
however, drives the $\rm HNC + H \rightarrow HCN +H$ reaction that has
a temperature barrier of $T\sim 200$~K, and HCN has an abundance that
is two orders of magnitude higher than HNC. In the low density case
(2) both the HCN and HNC abundances, which are similar, decrease fast
with increasing cosmic ray rates. This is a direct results of the
decrease of H$_2$ and H$_3^+$, which are important in the formation of
these molecules.

\textbf{\textit{HCO$^+$ and CS (Fig. \ref{Model_abun_HCOp_CS}):}} For
Model sets 1a and 1b, the HCO$^+$ abundances are enhanced with
increasing CR rates. The formation occurs through $\rm H_3^+ + CO
\rightarrow HCO^+ + H_2$, which is fast at all temperatures, and is
aided by $\rm CH + O \rightarrow HCO^+ + e^-$ at high
temperatures. Destruction happens through recombination with electrons
but also through reactions with molecules such as H$_2$O, OH, and HCN.
In Model set 1b, the abundances of these species are especially
enhanced for the lower ( $\zeta=5\cdot 10^{-17}$ and $5\cdot
10^{-16}$~s$^{-1}$) cosmic ray rates as discussed in the previous
paragraphs, and suppress the HCO$^+$ abundance compared to the models
without mechanical heating. In the low density models (2), where the
full transition to H$_2$ does not occur, the H$_3^+$ abundance is
automatically suppressed as well, giving smaller abundances with
increasing cosmic ray rates.

The sulfur chemistry contains many reactions with activation
barriers. CS has two formation channels, the first through $\rm S + CH
\rightarrow CS + H$ and the second through $\rm S + H_2 \rightarrow HS
+ H$ (with an energy barrier of $T\sim 6000$~K) followed by $\rm HS +
C \rightarrow CS + H$. CS is most abundant at column densities
$N_{\rm H} \gtrsim 5\cdot 10^{21}$~cm$^{-2}$, where temperatures are
not such that the second channel is very effective, although not
insignificant in Model set 2. The formation of CH on the other hand is
dependent on H$_3^+$, of which the abundance is highly dependent on CR
rate. Destruction happens through ion-molecule reactions with, e.g,
He$^+$, H$^+$, H$_3^+$, that are also dependent on CR rate. The
abundance is a very complex interplay between temperature and
ionization rates \citep[cf.,][]{Leen1988,Meijerink2008}. The
abundance of CS increases with CR rate in Model set 1a and 1b, while
Model set 1c shows a drop. Molecule formation becomes more and more
ineffective in the low density case, and the CS abundance decreases
significantly with CR rates, similar to, e.g., HCN and HNC.

\section{Integrated column densities}\label{Col_dens}

\begin{figure*}
  \centering
  \includegraphics[width=16cm,clip]{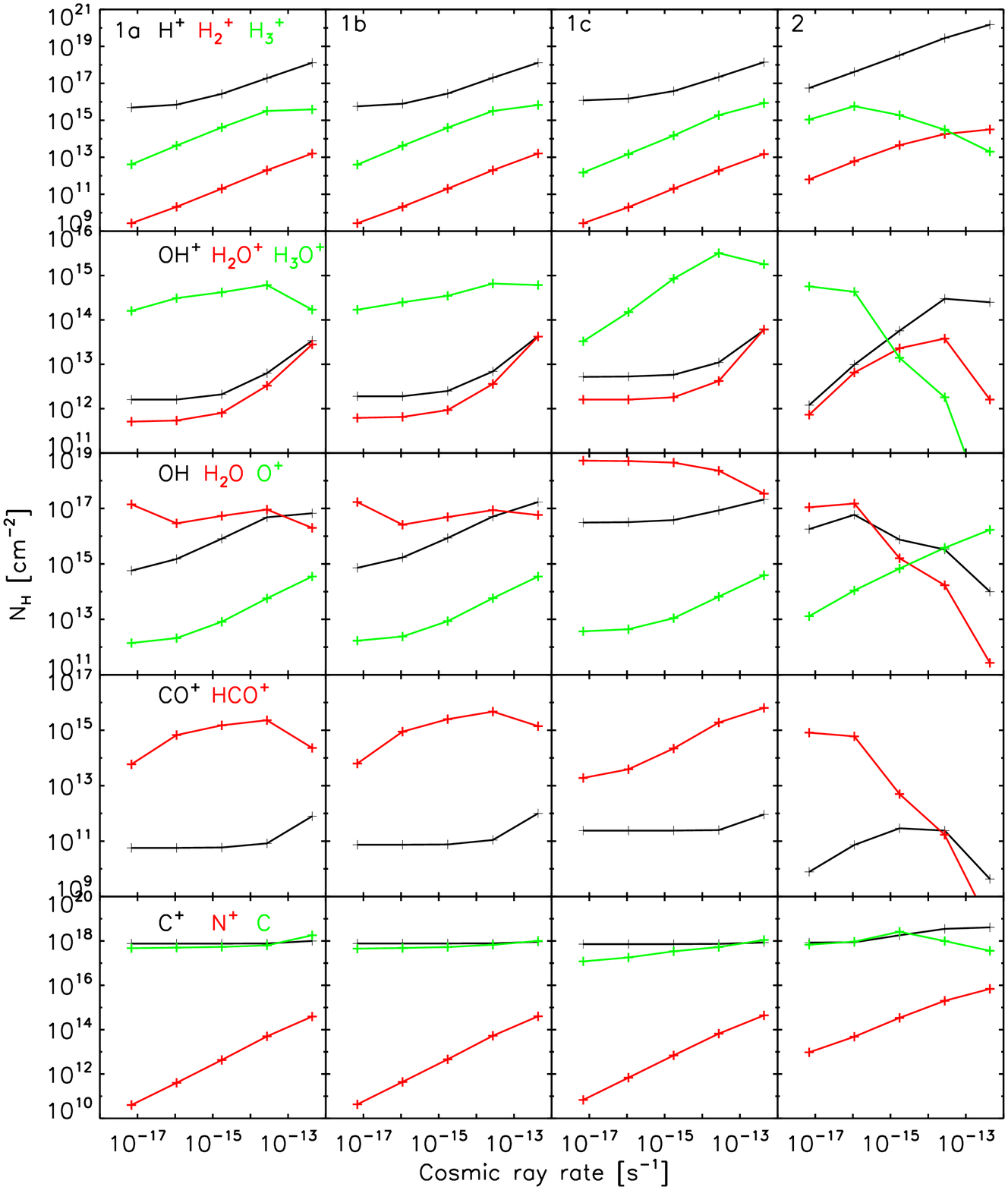}
  \caption{Column densities of diagnostic species.}
  \label{column_densities_water_species}
\end{figure*}

\begin{table}
\caption{Column densities for Model set 1a}   
\label{model_1a_column}    
\centering         
\begin{tabular}{l c c c c c}    
\hline\hline                
Species       & \multicolumn{5}{c}{Cosmic ray rate (s$^{-1}$)}  \\  
           & 5(-17) & 5(-16) & 5(-15) & 5(-14) & 5(-13) \\
\hline                     
\multicolumn{6}{c}{Column densities (cm$^{-2}$)} \\
\hline
H          & 4.2(21) & 4.2(21) & 4.3(21) & 5.2(21) & 1.2(22) \\
H$_2$      & 1.4(22) & 1.4(22) & 1.4(22) & 1.4(22) & 1.0(22) \\
H$^+$      & 4.9(15) & 7.0(15) & 2.7(16) & 1.9(17) & 1.3(18) \\
H$_2^+$    & 2.7(09) & 2.1(10) & 2.0(11) & 2.0(12) & 1.6(13) \\
H$_3^+$    & 4.1(12) & 4.3(13) & 4.1(14) & 3.2(15) & 3.9(15) \\
e$^-$      & 8.1(17) & 8.1(17) & 8.4(17) & 1.0(18) & 2.5(18) \\
CH$^+$     & 2.8(12) & 2.7(12) & 2.7(12) & 2.4(12) & 2.0(12) \\
CH$_2^+$   & 8.7(11) & 8.7(11) & 8.7(11) & 8.4(11) & 9.4(11) \\
CH         & 3.1(13) & 3.3(13) & 5.2(13) & 1.1(14) & 8.7(13) \\
O          & 6.6(18) & 6.4(18) & 6.4(18) & 6.8(18) & 9.3(18) \\
O$^+$      & 1.4(12) & 2.1(12) & 8.2(12) & 5.7(13) & 3.5(14) \\
OH$^+$     & 1.6(12) & 1.6(12) & 2.1(12) & 6.3(12) & 3.4(13) \\
H$_2$O$^+$ & 5.1(11) & 5.4(11) & 8.0(11) & 3.3(12) & 2.8(13) \\
H$_3$O$^+$ & 1.6(14) & 3.1(14) & 4.2(14) & 6.1(14) & 1.7(14) \\
OH         & 5.7(14) & 1.5(15) & 8.1(15) & 4.8(16) & 6.7(16) \\
H$_2$O     & 1.4(17) & 2.9(16) & 5.4(16) & 9.1(16) & 2.0(16) \\
C$^+$      & 7.6(17) & 7.6(17) & 7.6(17) & 7.7(17) & 1.0(18) \\
C          & 4.7(17) & 5.0(17) & 5.4(17) & 6.3(17) & 1.8(18) \\
CO         & 3.2(18) & 3.2(18) & 3.2(18) & 3.1(18) & 1.7(18) \\
CO$^+$     & 5.7(10) & 5.7(10) & 5.9(10) & 8.3(10) & 7.9(11) \\
HCO$^+$    & 5.9(13) & 6.7(14) & 1.5(15) & 2.3(15) & 2.3(14) \\
N$^+$      & 4.0(10) & 4.0(11) & 4.3(12) & 5.0(13) & 3.9(14) \\
NH$^+$     & 3.5(11) & 3.5(11) & 3.3(11) & 2.9(11) & 3.9(11) \\
NH$_2^+$   & 1.3(11) & 1.3(11) & 1.4(11) & 2.3(11) & 5.7(11) \\
HCN        & 4.3(13) & 5.7(13) & 5.6(13) & 1.2(14) & 3.2(13) \\
HNC        & 2.9(13) & 3.5(13) & 7.6(13) & 1.8(14) & 5.0(13) \\
CS         & 3.1(13) & 8.4(13) & 2.5(14) & 1.6(14) & 3.9(12) \\
\hline                           
\multicolumn{6}{l}{Note - Numbers in parentheses: $a(b) = a\cdot 10^{b}$.} 
\end{tabular}
\end{table}



\begin{table}
\caption{Column densities for Model set 1b}   
\label{model_1b_column}    
\centering         
\begin{tabular}{l c c c c c}    
\hline\hline                
Species       & \multicolumn{5}{c}{Cosmic ray rate (s$^{-1}$)}  \\  
           & 5(-17) & 5(-16) & 5(-15) & 5(-14) & 5(-13) \\
\hline                     
\multicolumn{6}{c}{Column densities (cm$^{-2}$)} \\
\hline
H          & 4.2(21) & 4.3(21) & 4.3(21) & 5.1(21) & 1.3(22) \\
H$_2$      & 1.4(22) & 1.4(22) & 1.4(22) & 1.4(22) & 1.0(22) \\
H$^+$      & 5.7(15) & 7.9(15) & 2.8(16) & 2.0(17) & 1.3(18) \\
H$_2^+$    & 2.7(09) & 2.1(10) & 2.0(11) & 2.0(12) & 1.6(13) \\
H$_3^+$    & 4.0(12) & 4.2(13) & 4.0(14) & 3.2(15) & 6.7(15) \\
e$^-$      & 8.2(17) & 8.2(17) & 8.5(17) & 1.0(18) & 2.3(18) \\
CH$^+$     & 3.2(12) & 3.2(12) & 3.1(12) & 2.9(12) & 2.3(12) \\
CH$_2^+$   & 1.0(12) & 1.0(12) & 1.0(12) & 1.0(12) & 9.8(11) \\
CH         & 3.1(13) & 3.3(13) & 7.0(13) & 1.6(14) & 5.9(13) \\
O          & 6.6(18) & 6.6(18) & 6.7(18) & 7.0(18) & 8.1(18) \\
O$^+$      & 1.7(12) & 2.4(12) & 8.6(12) & 5.8(13) & 3.5(14) \\
OH$^+$     & 1.9(12) & 1.9(12) & 2.5(12) & 6.9(12) & 4.3(13) \\
H$_2$O$^+$ & 6.2(11) & 6.5(11) & 9.3(11) & 3.6(12) & 4.2(13) \\
H$_3$O$^+$ & 1.7(14) & 2.5(14) & 3.5(14) & 6.6(14) & 6.1(14) \\
OH         & 7.2(14) & 1.7(15) & 8.6(15) & 5.1(16) & 1.7(17) \\
H$_2$O     & 1.7(17) & 2.6(16) & 4.9(16) & 8.7(16) & 5.8(16) \\
C$^+$      & 7.7(17) & 7.7(17) & 7.7(17) & 7.8(17) & 9.1(17) \\
C          & 4.5(17) & 4.8(17) & 5.3(17) & 6.7(17) & 1.0(18) \\
CO         & 3.2(18) & 3.2(18) & 3.2(18) & 3.0(18) & 2.5(18) \\
CO$^+$     & 7.4(10) & 7.4(10) & 7.6(10) & 1.1(11) & 1.0(12) \\
HCO$^+$    & 6.3(13) & 8.9(14) & 2.5(15) & 4.7(15) & 1.4(15) \\
N$^+$      & 4.3(10) & 4.4(11) & 4.6(12) & 5.3(13) & 4.0(14) \\
NH$^+$     & 3.1(11) & 3.1(11) & 2.9(11) & 2.5(11) & 3.8(11) \\
NH$_2^+$   & 1.1(11) & 1.1(11) & 1.2(11) & 2.0(11) & 7.3(11) \\
HCN        & 1.2(14) & 2.8(14) & 8.0(14) & 9.5(14) & 9.0(13) \\
HNC        & 7.5(13) & 1.1(14) & 2.5(14) & 3.1(14) & 8.9(13) \\
CS         & 8.6(11) & 5.2(12) & 2.3(13) & 5.0(13) & 1.0(13) \\
\hline                           
\end{tabular}
\end{table}

\begin{table}
\caption{Column densities for Model set 1c}   
\label{model_1c_column}    
\centering         
\begin{tabular}{l c c c c c}    
\hline\hline                
Species       & \multicolumn{5}{c}{Cosmic ray rate (s$^{-1}$)}  \\  
           & 5(-17) & 5(-16) & 5(-15) & 5(-14) & 5(-13) \\
\hline                     
\multicolumn{6}{c}{Column densities (cm$^{-2}$)} \\
\hline
H          & 5.1(21) & 5.1(21) & 5.3(21) & 6.2(21) & 1.4(22) \\
H$_2$      & 1.4(22) & 1.4(22) & 1.4(22) & 1.3(22) & 9.3(21) \\
H$^+$      & 1.2(16) & 1.5(16) & 3.8(16) & 2.2(17) & 1.4(18) \\
H$_2^+$    & 2.7(09) & 2.0(10) & 2.0(11) & 1.9(12) & 1.5(13) \\
H$_3^+$    & 1.5(12) & 1.5(13) & 1.5(14) & 1.9(15) & 8.6(15) \\
e$^-$      & 7.8(17) & 7.8(17) & 8.1(17) & 1.0(18) & 2.3(18) \\
CH$^+$     & 5.7(12) & 5.7(12) & 5.6(12) & 5.5(12) & 5.2(12) \\
CH$_2^+$   & 1.9(12) & 1.9(12) & 1.9(12) & 1.9(12) & 2.1(12) \\
CH         & 2.9(13) & 3.8(13) & 4.5(13) & 7.4(13) & 6.2(13) \\
O          & 3.2(18) & 3.2(18) & 3.4(18) & 5.1(18) & 7.8(18) \\
O$^+$      & 3.7(12) & 4.4(12) & 1.1(13) & 6.6(13) & 3.9(14) \\
OH$^+$     & 5.2(12) & 5.3(12) & 5.8(12) & 1.1(13) & 5.9(13) \\
H$_2$O$^+$ & 1.6(12) & 1.6(12) & 1.8(12) & 4.2(12) & 6.1(13) \\
H$_3$O$^+$ & 3.3(13) & 1.5(14) & 8.5(14) & 3.2(15) & 1.8(15) \\
OH         & 3.1(16) & 3.2(16) & 3.8(16) & 8.5(16) & 2.1(17) \\
H$_2$O     & 5.3(18) & 5.1(18) & 4.5(18) & 2.3(18) & 3.4(17) \\
C$^+$      & 7.2(17) & 7.2(17) & 7.2(17) & 7.4(17) & 8.5(17) \\
C          & 1.2(17) & 1.8(17) & 3.4(17) & 5.3(17) & 1.1(18) \\
CO         & 2.3(18) & 2.5(18) & 2.9(18) & 3.1(18) & 2.5(18) \\
CO$^+$     & 2.4(11) & 2.4(11) & 2.4(11) & 2.5(11) & 9.2(11) \\
HCO$^+$    & 1.9(13) & 3.9(13) & 2.2(14) & 1.9(15) & 6.4(15) \\
N$^+$      & 6.8(10) & 6.8(11) & 6.9(12) & 6.6(13) & 4.4(14) \\
NH$^+$     & 1.4(11) & 1.3(11) & 1.3(11) & 1.2(11) & 2.7(11) \\
NH$_2^+$   & 3.5(10) & 3.5(10) & 4.1(10) & 1.0(11) & 6.1(11) \\
HCN        & 8.0(17) & 8.2(17) & 4.4(17) & 9.5(16) & 3.5(15) \\
HNC        & 1.1(16) & 8.0(15) & 5.5(15) & 1.2(15) & 8.0(13) \\
CS         & 4.6(14) & 3.2(14) & 6.3(13) & 3.5(13) & 8.9(12) \\
\hline                           
\end{tabular}
\end{table}

\begin{table}
\caption{Column densities for Model set 2}
\label{model_2_column}    
\centering         
\begin{tabular}{l c c c c c}    
\hline\hline                
Species       & \multicolumn{5}{c}{Cosmic ray rate (s$^{-1}$)}  \\  
           & 5(-17) & 5(-16) & 5(-15) & 5(-14) & 5(-13) \\
\hline                     
\multicolumn{6}{c}{Column densities (cm$^{-2}$)} \\
\hline
H          & 4.4(21) & 6.6(21) & 1.6(22) & 2.7(22) & 3.2(22) \\
H$_2$      & 1.4(22) & 1.3(22) & 8.6(21) & 2.8(21) & 5.2(20) \\
H$^+$      & 5.6(16) & 4.2(17) & 3.3(18) & 2.8(19) & 1.5(20) \\
H$_2^+$    & 6.3(11) & 6.0(12) & 4.5(13) & 1.8(14) & 3.2(14) \\
H$_3^+$    & 1.1(15) & 5.8(15) & 1.9(15) & 3.1(14) & 2.0(13) \\
e$^-$      & 9.5(17) & 1.4(18) & 5.5(18) & 3.4(19) & 1.7(20) \\
CH$^+$     & 7.9(10) & 4.1(11) & 5.8(11) & 1.3(11) & 5.7(10) \\
CH$_2^+$   & 1.4(11) & 3.9(11) & 6.5(11) & 1.3(11) & 6.1(09) \\
CH         & 4.5(13) & 1.7(14) & 6.0(13) & 4.3(12) & 4.7(11) \\
O          & 6.8(18) & 7.8(18) & 1.1(19) & 1.1(19) & 1.1(19) \\
O$^+$      & 1.3(13) & 1.1(14) & 6.8(14) & 3.8(15) & 1.7(16) \\
OH$^+$     & 1.2(12) & 9.8(12) & 5.7(13) & 3.0(14) & 2.5(14) \\
H$_2$O$^+$ & 7.3(11) & 6.5(12) & 2.3(13) & 3.8(13) & 1.6(12) \\
H$_3$O$^+$ & 5.7(14) & 4.3(14) & 1.4(13) & 1.8(12) & 3.0(09) \\
OH         & 1.8(16) & 5.9(16) & 7.5(15) & 3.3(15) & 1.0(14) \\
H$_2$O     & 1.1(17) & 1.5(17) & 1.6(15) & 1.7(14) & 2.7(11) \\
C$^+$      & 8.4(17) & 8.7(17) & 1.8(18) & 3.5(18) & 4.1(18) \\
C          & 6.8(17) & 9.3(17) & 2.6(18) & 9.9(17) & 3.6(17) \\
CO         & 2.9(18) & 2.7(18) & 4.0(16) & 1.8(15) & 3.9(12) \\
CO$^+$     & 7.8(09) & 7.3(10) & 2.9(11) & 2.4(11) & 4.3(09) \\
HCO$^+$    & 8.2(14) & 6.0(14) & 5.0(12) & 1.7(11) & 9.7(07) \\
N$^+$      & 9.5(12) & 4.8(13) & 3.4(14) & 2.0(15) & 6.9(15) \\
NH$^+$     & 1.8(11) & 1.8(11) & 2.2(11) & 1.3(12) & 4.8(12) \\
NH$_2^+$   & 2.6(10) & 1.6(11) & 2.3(11) & 3.0(11) & 4.9(10) \\
HCN        & 2.2(14) & 1.2(14) & 2.6(12) & 1.5(10) & 3.5(06) \\
HNC        & 2.5(14) & 1.8(14) & 6.1(12) & 3.4(10) & 3.3(06) \\
CS         & 1.8(14) & 6.2(13) & 2.5(10) & 4.9(06) & 3.2(03) \\
\hline                           
\end{tabular}
\end{table}

\subsection{Trends}

The integrated column densities for the total cloud ($N_{\rm H}=3\cdot
10^{22}$~cm$^{-2}$) are shown in Tables \ref{model_1a_column} through
\ref{model_2_column} for the species discussed in the previous
section. The main results are summarized in
Fig. \ref{column_densities_water_species}.

\textbf{\textit{Hydrogen chemistry (top row
  Fig. \ref{column_densities_water_species}):}} The integrated column
densities of atomic and molecular hydrogen vary over a factor of ~3
(10) for the high (low) density models over the whole range of cosmic
ray rates. The ionic species of hydrogen, H$^+$, H$_2^+$, and H$_3^+$
show much larger variations. The column densities of $N({\rm H}^+)$
vary from $\sim 10^{16}-10^{18}$~cm$^{-2}$
($10^{17}-10^{20}$~cm$^{-2}$) for the high (low density
models). H$_3^+$ increases (decreases) 4 (2) orders of magnitude in
the high (low) density case.

\textbf{\textit{Water chemistry (row 2 and 3 of
    Fig. \ref{column_densities_water_species}):}} The O$^+$ column
densities show a nice correlation with CR rate, and increases 2 to 3
orders of magnitude over the range of considered CR rates. Although
there are large variations in absolute column densities for OH and
H$_2$O between the different models, the OH/H$_2$O ratio increases
with cosmic ray rate, from $N({\rm OH})/N({\rm H_2O})\sim 4\cdot
10^{-3}$ ($\sim 0.2$) to $\sim 1-3$ ($\sim 400$) for the high (low)
density models.

The OH$^+$ and H$_2$O$^+$ show a strong response to the CR rate for
$\zeta\gtrsim 5\cdot 10^{-15}$~s$^{-1}$ in the high density
case. Above this threshold, the column density increases with a factor
$\sim 10 - 20$ for OH$^+$ and $\sim 50 - 60$ for H$_2$O$^+$. In the
low density case, the column densities for both species show a steep
rise with CR rate up to $\zeta\sim 5\cdot 10^{-14}$~s$^{-1}$, after
which it levels off for OH$^+$ and decreases for H$_2$O$^+$. The
trends of H$_3$O$^+$ generally show an increase (decrease) with CR
rate in the high (low) density case. The response is more direct when
high mechanical heating is added and temperatures are higher.

\textbf{\textit{CO$^+$ and HCO$^+$ (row 4 in
    Fig. \ref{column_densities_water_species}):}} HCO$^+$ does not
show an obvious trend with increasing CR rates in the high density
case. The HCO$^+$ abundance anti-correlates with CR rate in the low
density case. It does not seem to be a tracer that can be easily used
as a CR tracer, but it does provide information when used in
combination with e.g. the water chemistry. CO$^+$ is very irresponsive
to the CR rates at high density except for very high CR rates ($\zeta
> 5\cdot 10^{-14}$~s$^{-1}$). In the low density case, $N({\rm
  CO}^+)$ does not show a very obvious trend.

\textbf{\textit{C$^+$, N$^+$, and C (bottom row in
  Fig. \ref{column_densities_water_species}):}} Neutral and singly
ionized carbon have very constant column densities over the range of
CR rates considered. N$^+$ on the other hand correlates very well with
the CR rates, and therefore the N$^+$/C or N$^+$/C$^+$ column density
ratio can be diagnostic in estimating the CR rate.

\textbf{\textit{ Other species:}} Despite the fact that many species
show large {\it local} variations in a cloud when varying CR rates, it
does not always result in significant changes in the integrated column
densities in high density clouds. Examples are CH$^+$ and CH$_2^+$
which are the same within 5 percent over the whole range. In the low
density models, variations are larger, but the integrated column
densities are 1 or 2 orders of magnitude smaller. There are also
species that are not easy to interpret, such as NH$^+$ and
NH$_2^+$. The trends show a minimum or a maximum, when varying the CR
rates. A number of species we considered have a strong dependence on
temperature and ionization rate. Especially CS exhibits are very
complex interplay, and does not show a particular trend.

\subsection{Uncertainties}

Molecular hydrogen, H$_2$, is a key species in the chemistry of the
ISM. The formation of H$_2$, which mainly occurs on grain surfaces at
these densities and metallicity, depend on the specific properties and
temperatures of the grains, and is subject to a considerable
uncertainty \citep[cf.,][]{Cazaux2004}. As a result the H$_2$
abundance is uncertain as well, especially when the gas is not
predominantly molecular, i.e., in the radical region.

Therefore, the column densities of key species such as O$^+$, OH,
OH$^+$, and H$_2$O$^+$ might be affected, especially when the
integrated column density is dominated by the abundances in the
radical region. In order to investigate this, we lowered the H$_2$
formation rate by a factor 3, and here we summarize the results for
some of the species, considering Model set 1a:

\textbf{\textit{H$_2$, H$_2^+$, and H$^+$}}: The abundance of $H_2$ is
lower by a factor of $\sim 3$ in the unshielded region of the cloud,
and the H to H$_2$ transition occurs now more gradual. In the
molecular region, there is only a noticeable effect when there is no
full transition to H$_2$, which is for the highest considered CR rate,
$\zeta = 5\cdot 10^{-13}$~s$^{-1}$. The column density of atomic
hydrogen is $\sim10$ percent higher for the lowest up to $\sim80$
percent higher for the highest CR rate. Molecular hydrogen column
densities only change significantly for the highest two CR rates,
$\sim20$ and $\sim40$ percent lower, respectively.

The H$^+$ column density is lower by a factor $\sim3$ for the lowest
and $\sim 5$ to $\sim 15$ percent higher for the highest two CR
rates. The H$^+$ column density is dominated by H$_2^+$ + H
$\rightarrow$ H$^+$ + H$_2$ in the radical region at
$\zeta=5\cdot10^{-17}$~s$^{-1}$ and thus lower, while at higher CR
rates the column density is increased because of the only partial
transition to H$_2$

\textbf{\textit{O$^+$, OH, OH$^+$, H$_2$O and H$_2$O$^+$, and
    H$_3$O$^+$}}: The trend seen for the O$^+$ column density is very
close to what is seen for H$^+$, i.e, a factor of $\sim3$ lower for
$\zeta=5 \cdot10^{-17}$~s$^{-1}$, while slightly higher for the highest
CR rates. Similar trends are seen for OH$^+$ and H$_2$O$^+$.

H$_2$O is only significantly affected (lowered by a factor $\sim 2$)
at the highest CR rate. OH column density is lower by a factor $\sim3$
for the lowest CR rates, and dominated by the abundance in the
unshielded region of the cloud. This difference is much less (5-30
percent) for the higher CR rates. H$_2$O and H$_3$O$^+$ column
densities are only affected for the higest two CR rates. \\
 
In all we conclude that the overall trends are not affected by
uncertainties in the H$_2$ abundance, the main driver of the species
of interest in our paper. The largest changes are seen when the
integrated column density of a species is dominated by its abundance
in the radical region, i.e., in the models with small CR rates.

\section{Integrated line fluxes}\label{Obs_diag}

Line fluxes for a number of species discussed in the previous section
are shown in Tables \ref{fluxes_1a} through \ref{fluxes_2}. Here we
highlight the main results:

\textbf{\textit{Fine-structure lines:}} The two important cooling lines,
[OI]~63~$\mu$m and [CII]~158~$\mu$m, that mainly determine the energy
budget of interstellar clouds, have very similar fluxes for different
CR rates in the high density case. [OI]~63~$\mu$m is the same within a
factor of 2 in the high density case for all mechanical heating and CR
rates, but does increase a factor 5 in the low density case. The
critical density for excitation of the [OI]~63~$\mu$m is
$n_{crit}=5\cdot10^5$~cm$^{-3}$, and the low density model is not
even close to being thermalized, and temperature differences cause a
large change in emitted flux. The [CII]~158~$\mu$m has a very low
critical density $n_{crit}=2.8\cdot 10^{3}$~cm$^{-3}$, and the
increase in fluxes is not explained by the rise in temperature at
small column densities. Additional flux is emitted in low density
clouds, because there is no full transition to neutral carbon and CO,
for the highest CR rates (see Section 2). This also explains why the
two neutral carbon lines, [CI]~610~$\mu$m and [CI]~370~$\mu$m, show an
increase in the high density models. The low density models have a
maximum intensity at $\zeta\sim 5\cdot 10^{-15}$~s$^{-1}$, since for
higher CR rates the main carbon budget remains in C$^+$. The two
ionized nitrogen lines, [NII]~205~$\mu$m and [NII]~122~$\mu$m,
increase by 4 (3) orders of magnitude in the high (low) density
case. Especially low density clouds are expected to yield detectable
fluxes: the [NII]~122~$\mu$m lines has only a 10 times weaker flux
than the [CII] line at the highest CR rate. Unfortunately, O$^+$ does
not have fine-structure transitions in the mid and
far-infrared. Available transitions are in the visible and UV, and
have very high excitation energies.

\textbf{\textit{H$_3^+$~95~$\mu$m:}} This transition is calcalated
following the discussion in \citet{Pan1986}. The fluxes in the high
density case are very similar to the [NII]~122~$\mu$m, not extremely
strong, but potentially detectable, when clouds are exposed to very
large CR fluxes.

\textbf{\textit{HCN, HNC, and HCO$^+$:}} The HCN and HNC do not seem
very helpful in the discussion of cosmic ray rates, as there are no
obvious trends. The HCN/HNC ratio does show a response to mechanical
heating as already pointed out in \citet{Loenen2008}, and especially
HCN is boosted by a few orders of magnitude. An exception is the high
density with the high CR rate, where He$^+$ destroys the HCN and HNC
very effectively. Also HCO$^+$ does not exhibit very obvious trends,
except for very high mechanical heating rates (model set 1c).

\textbf{\textit{CO and $^{13}$CO:}} In the high density models, there
is very little response when increasing the cosmic ray flux. Fluxes
are boosted, when mechanical heating plays an important role in the
energy budget of the gas, but, contrary to the HCN and HNC lines, it
only affects high-$J$ transitions. When $^{13}$CO line fluxes are also
available, mechanical heating effects are potentially also seen in the
lower-$J$ transitions ($J\lesssim 3$). The CO to $^{13}$CO line ratio
increases, indicating smaller optical depths, while CO fluxes only
decrease by a factor 2

\begin{table}
\caption{Line intensities for Model set 1a}   
\label{fluxes_1a}    
\centering         
\begin{tabular}{l c c c c c}    
\hline\hline                
Line       & \multicolumn{5}{c}{Cosmic ray rate (s$^{-1}$)}  \\  
           & 5(-17) & 5(-16) & 5(-15) & 5(-14) & 5(-13) \\
\hline                     
\multicolumn{6}{c}{Line intensities (erg~s$^{-1}$~cm$^{-2}$~sr$^{-1}$)} \\
\hline
$[$CII$]$ 158~$\mu$m & 1.1(-3) & 1.1(-3) & 1.1(-3) & 1.1(-3) & 1.3(-3) \\
$[$CI$]$ 610~$\mu$m  & 3.7(-6) & 3.9(-6) & 4.2(-6) & 4.8(-6) & 1.1(-5) \\
$[$CI$]$ 370~$\mu$m  & 2.1(-5) & 2.2(-5) & 2.3(-5) & 2.6(-5) & 6.2(-5) \\
$[$OI$]$ 63~$\mu$m   & 7.8(-2) & 7.8(-2) & 7.8(-2) & 8.2(-2) & 9.2(-2) \\ 
$[$OI$]$ 146~$\mu$m  & 2.3(-3) & 2.3(-3) & 2.3(-3) & 2.4(-3) & 2.9(-3) \\ 
$[$NII$]$ 205~$\mu$m & 4.4(-11) & 4.4(-10) & 4.7(-9) & 5.5(-8) & 4.2(-7) \\ 
$[$NII$]$ 122~$\mu$m & 6.6(-10) & 6.6(-9) & 7.1(-8) & 8.4(-7) & 6.6(-6) \\
H$_3^+$ 95~$\mu$m & 7.0(-10) & 6.2(-9) & 6.0(-8) & 6.5(-7) & 1.9(-6) \\
HCN $J=  1-0$  & 1.5(-8) & 1.8(-8) & 1.8(-8) & 2.9(-8) & 1.1(-8) \\
HCN $J=  2-1$  & 1.3(-7) & 1.5(-7) & 1.5(-7) & 2.2(-7) & 1.3(-7) \\
HCN $J=  3-2$  & 3.2(-7) & 3.7(-7) & 3.6(-7) & 5.8(-7) & 3.3(-7) \\
HCN $J=  4-3$  & 3.8(-7) & 4.7(-7) & 4.5(-7) & 8.9(-7) & 4.4(-7) \\
HCN $J=  5-4$  & 3.3(-7) & 4.0(-7) & 3.9(-7) & 7.7(-7) & 4.0(-7) \\
HCN $J=  6-5$  & 2.6(-7) & 3.0(-7) & 3.0(-7) & 5.2(-7) & 3.1(-7) \\
HCN $J=  7-6$  & 2.0(-7) & 2.2(-7) & 2.2(-7) & 3.2(-7) & 2.2(-7) \\
HNC $J=  1- 0$ & 1.1(-8) &1.3(-8) & 2.0(-8) & 3.6(-8) & 1.7(-8) \\
HNC $J=  2- 1$ & 9.9(-8) &1.1(-7) & 1.6(-7) & 2.5(-7) & 1.7(-7)\\
HNC $J=  3- 2$ & 2.1(-7) & 2.4(-7) & 3.9(-7) & 6.7(-7) & 4.5(-7) \\
HNC $J=  4- 3$ & 2.3(-7) & 2.6(-7) & 4.9(-7) & 1.1(-6) & 6.3(-7) \\
HNC $J=  5- 4$ & 1.9(-7) & 2.0(-7) & 3.4(-7) & 1.1(-6) & 5.7(-7) \\
HNC $J=  6- 5$ & 1.4(-7) & 1.5(-7) & 2.2(-7) & 6.1(-7) & 4.2(-7) \\
HNC $J=  7- 6$ & 9.7(-8) & 1.0(-7) & 1.3(-7) & 3.3(-7) & 2.8(-7) \\
HCO$^+$ $J=  1- 0$ & 2.7(-8) & 1.0(-7) & 1.5(-7) & 1.6(-7) & 5.5(-8) \\
HCO$^+$ $J=  2- 1$ & 2.8(-7) & 7.7(-7) & 9.8(-7) & 1.2(-6) & 7.3(-7) \\
HCO$^+$ $J=  3- 2$ & 8.1(-7) & 2.2(-6) & 2.7(-6) & 3.3(-6) & 3.1(-6) \\
HCO$^+$ $J=  4- 3$ & 1.4(-6) & 4.4(-6) & 5.5(-6) & 6.8(-6) & 6.8(-6) \\
HCO$^+$ $J=  5- 4$ & 1.6(-6) & 7.2(-6) & 9.4(-6) & 1.2(-5) & 9.4(-6) \\
HCO$^+$ $J=  6- 5$ & 1.2(-6) & 1.0(-5) & 1.4(-5) & 1.9(-5) & 1.3(-5) \\
HCO$^+$ $J=  7- 6$ & 8.2(-7) & 1.2(-5) & 2.0(-5) & 2.8(-5) & 1.5(-5) \\
CO $J=  1- 0$ & 3.0(-7) & 3.0(-7) & 3.1(-7) & 3.2(-7) & 3.1(-7) \\
CO $J=  2- 1$ & 2.9(-6) & 3.0(-6) & 3.0(-6) & 3.3(-6) & 3.9(-6) \\
CO $J=  3- 2$ & 1.0(-5) & 1.0(-5) & 1.1(-5) & 1.2(-5) & 1.5(-5) \\
CO $J=  4- 3$ & 2.4(-5) & 2.4(-5) & 2.5(-5) & 2.8(-5) & 3.9(-5) \\
CO $J=  5- 4$ & 4.3(-5) & 4.4(-5) & 4.6(-5) & 5.3(-5) & 7.8(-5)\\
CO $J=  6- 5$ & 6.7(-5) & 6.8(-5) & 7.1(-5) & 8.4(-5) & 1.3(-4)\\
CO $J=  7- 6$ & 9.2(-5) & 9.4(-5) & 9.8(-5) & 1.2(-4) & 2.0(-4)\\
CO $J=  8- 7$ & 1.1(-4) & 1.1(-4) & 1.2(-4) & 1.5(-4) & 2.8(-4) \\
CO $J=  9- 8$ & 1.2(-4) & 1.3(-4) & 1.3(-4) & 1.7(-4) & 3.6(-4) \\
CO $J= 10- 9$ & 1.1(-4) & 1.2(-4) & 1.2(-4) & 1.8(-4) & 4.3(-4) \\
CO $J= 11-10$ & 8.2(-5) & 8.9(-5) & 9.4(-5) & 1.6(-4) & 4.8(-4) \\
CO $J= 12-11$ & 3.8(-5) & 4.3(-5) & 4.8(-5) & 1.1(-4) & 5.0(-4) \\
CO $J= 13-12$ & 1.2(-5) & 1.4(-5) & 1.6(-5) & 5.1(-5) & 4.7(-4) \\
CO $J= 14-13$ & 4.7(-6) & 5.2(-6) & 5.7(-6) & 1.8(-5) & 3.8(-4) \\
CO $J= 17-16$ & 3.4(-6) & 3.4(-6) & 3.4(-6) & 3.8(-6) & 5.8(-5) \\
CO $J= 20-19$ & 3.6(-6) & 3.6(-6) & 3.6(-6) & 3.5(-6) & 7.8(-6) \\
CO $J= 24-23$ & 2.4(-6) & 2.4(-6) & 2.3(-6) & 2.2(-6) & 2.4(-6) \\
CO $J= 30-29$ & 4.9(-7) & 4.9(-7) & 4.8(-7) &4.4(-7) & 4.9(-7) \\
CO $J= 35-34$ & 1.0(-7) & 1.0(-7) & 1.0(-7) & 9.9(-8) & 1.3(-7) \\
$^{13}$CO $J=  1- 0$ & 4.1(-8) & 4.1(-8) & 4.0(-8) & 3.6(-8) & 1.4(-8) \\
$^{13}$CO $J=  2- 1$ & 7.9(-7) & 7.9(-7) & 7.9(-7) & 7.6(-7) & 3.7(-7) \\
$^{13}$CO $J=  3- 2$ & 3.4(-6) & 3.4(-6) & 3.4(-6) & 3.5(-6) & 2.2(-6) \\
$^{13}$CO $J=  4- 3$ & 8.4(-6) & 8.5(-6) & 8.5(-6) & 9.1(-6) & 7.0(-6) \\
$^{13}$CO $J=  5- 4$ & 1.5(-5) & 1.5(-5) & 1.5(-5) & 1.7(-5) & 1.5(-5) \\
$^{13}$CO $J=  6- 5$ & 2.2(-5) & 2.2(-5) & 2.2(-5) & 2.6(-5) & 2.7(-5) \\
$^{13}$CO $J=  7- 6$ & 2.5(-5) & 2.6(-5) & 2.6(-5) & 3.3(-5) & 3.9(-5) \\
$^{13}$CO $J=  8- 7$ & 2.3(-5) & 2.4(-5) & 2.5(-5) & 3.4(-5) & 5.0(-5) \\
$^{13}$CO $J=  9- 8$ & 1.6(-5) & 1.7(-5) & 1.8(-5) & 2.8(-5) & 5.6(-5) \\
$^{13}$CO $J= 10- 9$ & 8.3(-6) & 9.1(-6) & 9.7(-6) & 1.8(-5) & 5.5(-5) \\
$^{13}$CO $J= 11-10$ & 3.4(-6) & 3.8(-6) & 4.2(-6) & 9.6(-6) & 4.7(-5) \\
$^{13}$CO $J= 12-11$ & 1.2(-6) & 1.4(-6) & 1.5(-6) & 4.2(-6) & 3.6(-5) \\
$^{13}$CO $J= 13-12$ & 4.0(-7) & 4.4(-7) & 5.0(-7) & 1.6(-6) & 2.4(-5) \\
$^{13}$CO $J= 14-13$ & 1.5(-7) & 1.6(-7) & 1.8(-7) & 5.7(-7) & 1.4(-5) \\
\hline                           
\end{tabular}
\end{table}

\begin{table}
\caption{Line intensities for Model set 1b}   
\label{fluxes_1b}    
\centering         
\begin{tabular}{l c c c c c}    
\hline\hline                
Line       & \multicolumn{5}{c}{Cosmic ray rate (s$^{-1}$)}  \\  
           & 5(-17) & 5(-16) & 5(-15) & 5(-14) & 5(-13) \\
\hline                     
\multicolumn{6}{c}{Line intensities (erg~s$^{-1}$~cm$^{-2}$~sr$^{-1}$)} \\
\hline
$[$CII$]$ 158~$\mu$m & 1.1(-3) & 1.1(-3) & 1.1(-3) & 1.2(-3) & 1.3(-3) \\
$[$CI$]$ 610~$\mu$m  & 3.4(-6) & 3.6(-6) & 3.9(-6) & 4.9(-6) & 7.3(-6) \\ 
$[$CI$]$ 370~$\mu$m  & 2.4(-5) & 2.5(-5) & 2.8(-5) & 3.4(-5) & 5.2(-5) \\
$[$OI$]$ 63~$\mu$m   & 9.3(-2) & 9.4(-2) & 9.4(-2) & 9.8(-2) & 1.1(-1) \\
$[$OI$]$ 146~$\mu$m  & 3.7(-3) & 3.7(-3) & 3.8(-3) & 3.9(-3) & 4.1(-3) \\ 
$[$NII$]$ 205~$\mu$m & 4.7(-11) & 4.7(-10) & 5.0(-9) & 5.7(-8) & 4.3(-7) \\      
$[$NII$]$ 122~$\mu$m & 7.2(-10) & 7.2(-9) & 7.7(-8) & 8.8(-7) & 6.7(-6) \\
H$_3^+$ 95~$\mu$m & 3.6(-9) & 3.8(-8) & 3.6(-7) & 2.8(-6) & 6.0(-6) \\
HCN $J=  1- 0$ & 3.1(-8) & 5.8(-8) & 1.2(-7) & 1.3(-7) & 2.0(-8) \\
HCN $J=  2- 1$ & 2.9(-7) & 4.3(-7) & 6.7(-7) & 7.4(-7) & 2.8(-7) \\
HCN $J=  3- 2$ & 8.3(-7) & 1.2(-6) & 1.9(-6) & 2.0(-6) & 8.1(-7) \\
HCN $J=  4- 3$ & 1.6(-6) & 2.6(-6) & 4.1(-6) & 4.4(-6) & 1.5(-6) \\
HCN $J=  5- 4$ & 2.0(-6) & 4.2(-6) & 7.4(-6) & 8.1(-6) & 2.0(-6) \\
HCN $J=  6- 5$ & 1.8(-6) & 5.2(-6) & 1.2(-5) & 1.3(-5) & 1.9(-6) \\
HCN $J=  7- 6$ & 1.5(-6) & 4.6(-6) & 1.5(-5) & 1.8(-5) & 1.6(-6) \\
HNC $J=  1- 0$ & 2.2(-8) & 3.0(-8) & 5.5(-8) & 6.0(-8) & 2.2(-8) \\
HNC $J=  2- 1$ & 2.3(-7) & 2.9(-7) & 4.3(-7) & 4.7(-7) & 2.8(-7) \\
HNC $J=  3- 2$ & 6.5(-7) & 8.2(-7) & 1.2(-6) & 1.3(-6) & 8.1(-7) \\
HNC $J=  4- 3$ & 1.1(-6) & 1.5(-6) & 2.5(-6) & 2.8(-6) & 1.5(-6) \\
HNC $J=  5- 4$ & 1.2(-6) & 2.0(-6) & 4.0(-6) & 4.5(-6) & 1.9(-6) \\
HNC $J=  6- 5$ & 1.1(-6) & 1.8(-6) & 4.7(-6) & 5.6(-6) & 1.8(-6) \\
HNC $J=  7- 6$ & 9.2(-7) & 1.5(-6) & 3.8(-6) & 4.9(-6) & 1.5(-6) \\
HCO$^+$ $J=  1- 0$ & 1.9(-8) & 1.4(-7) & 3.3(-7) & 5.5(-7) & 1.9(-7) \\
HCO$^+$ $J=  2- 1$ & 3.8(-7) & 1.7(-6) & 2.9(-6) & 3.9(-6) & 2.5(-6) \\
HCO$^+$ $J=  3- 2$ & 1.3(-6) & 4.7(-6) & 6.5(-6) & 8.2(-6) & 6.6(-6) \\
HCO$^+$ $J=  4- 3$ & 2.7(-6) & 9.5(-6) & 1.3(-5) & 1.5(-5) & 1.3(-5) \\
HCO$^+$ $J=  5- 4$ & 4.1(-6) & 1.7(-5) & 2.3(-5) & 2.7(-5) & 2.3(-5) \\
HCO$^+$ $J=  6- 5$ & 4.9(-6) & 2.7(-5) & 3.8(-5) & 4.5(-5) & 3.7(-5) \\
HCO$^+$ $J=  7- 6$ & 4.5(-6) & 3.9(-5) & 5.9(-5) & 7.1(-5) & 5.5(-5) \\
CO $J=  1- 0$ & 4.2(-7) & 4.1(-7) & 4.1(-7) & 4.0(-7) & 3.6(-7) \\
CO $J=  2- 1$ & 6.1(-6) & 6.1(-6) & 6.2(-6) & 6.1(-6) & 5.8(-6) \\
CO $J=  3- 2$ & 2.5(-5) & 2.5(-5) & 2.5(-5) & 2.5(-5) & 2.4(-5) \\
CO $J=  4- 3$ & 6.3(-5) & 6.3(-5) & 6.4(-5) & 6.4(-5) & 6.4(-5) \\
CO $J=  5- 4$ & 1.3(-4) & 1.3(-4) & 1.3(-4) & 1.3(-4) & 1.3(-4) \\
CO $J=  6- 5$ & 2.2(-4) & 2.2(-4) & 2.3(-4) & 2.3(-4) & 2.3(-4) \\
CO $J=  7- 6$ & 3.5(-4) & 3.6(-4) & 3.6(-4) & 3.6(-4) & 3.7(-4) \\
CO $J=  8- 7$ & 5.2(-4) & 5.2(-4) & 5.3(-4) & 5.3(-4) & 5.5(-4) \\
CO $J=  9- 8$ & 7.1(-4) & 7.3(-4) & 7.3(-4) & 7.3(-4) & 7.7(-4) \\
CO $J= 10- 9$ & 9.4(-4) & 9.6(-4) & 9.7(-4) & 9.7(-4) & 1.0(-3) \\
CO $J= 11-10$ & 1.2(-3) & 1.2(-3) & 1.2(-3) & 1.2(-3) & 1.3(-3) \\
CO $J= 12-11$ & 1.5(-3) & 1.5(-3) & 1.5(-3) & 1.5(-3) & 1.6(-3) \\
CO $J= 13-12$ & 1.7(-3) & 1.8(-3) & 1.8(-3) & 1.7(-3) & 1.8(-3) \\
CO $J= 14-13$ & 2.0(-3) & 2.0(-3) & 2.0(-3) & 2.0(-3) & 2.1(-3) \\
CO $J= 17-16$ & 2.2(-3) & 2.4(-3) & 2.3(-3) & 2.2(-3) & 2.3(-3) \\
CO $J= 20-19$ & 9.1(-4) & 1.1(-3) & 1.0(-3) & 8.7(-4) & 1.2(-3) \\
CO $J= 24-23$ & 2.7(-5) & 3.7(-5) & 3.1(-5) & 2.6(-5) & 5.5(-5) \\
CO $J= 30-29$ & 8.0(-7) & 8.1(-7) & 7.8(-7) & 7.2(-7) & 7.8(-7) \\
CO $J= 35-34$ & 1.7(-7) & 1.7(-7) & 1.7(-7) & 1.6(-7) & 1.9(-7) \\
$^{13}$CO $J=  1- 0$ & 1.5(-8) & 1.4(-8) & 1.4(-8) & 1.4(-8) & 1.1(-8) \\
$^{13}$CO $J=  2- 1$ & 4.3(-7) & 4.2(-7) & 4.2(-7) & 4.0(-7) & 3.2(-7) \\
$^{13}$CO $J=  3- 2$ & 2.9(-6) & 2.8(-6) & 2.8(-6) & 2.7(-6) & 2.2(-6) \\
$^{13}$CO $J=  4- 3$ & 1.0(-5) & 1.0(-5) & 1.0(-5) & 9.6(-6) & 7.9(-6) \\
$^{13}$CO $J=  5- 4$ & 2.6(-5) & 2.5(-5) & 2.5(-5) & 2.4(-5) & 2.0(-5) \\
$^{13}$CO $J=  6- 5$ & 5.1(-5) & 5.0(-5) & 5.0(-5) & 4.8(-5) & 4.1(-5) \\
$^{13}$CO $J=  7- 6$ & 8.5(-5) & 8.5(-5) & 8.5(-5) & 8.1(-5) & 7.0(-5) \\
$^{13}$CO $J=  8- 7$ & 1.3(-4) & 1.3(-4) & 1.3(-4) & 1.2(-4) & 1.1(-4) \\
$^{13}$CO $J=  9- 8$ & 1.7(-4) & 1.7(-4) & 1.7(-4) & 1.7(-4) & 1.5(-4) \\
$^{13}$CO $J= 10- 9$ & 2.1(-4) & 2.1(-4) & 2.1(-4) & 2.0(-4) & 1.9(-4) \\
$^{13}$CO $J= 11-10$ & 2.4(-4) & 2.5(-4) & 2.4(-4) & 2.3(-4) & 2.2(-4) \\
$^{13}$CO $J= 12-11$ & 2.5(-4) & 2.6(-4) & 2.5(-4) & 2.4(-4) & 2.3(-4) \\
$^{13}$CO $J= 13-12$ & 2.4(-4) & 2.5(-4) & 2.4(-4) & 2.3(-4) & 2.4(-4) \\
$^{13}$CO $J= 14-13$ & 2.0(-4) & 2.2(-4) & 2.1(-4) & 2.0(-4) & 2.2(-4) \\
\hline                           
\end{tabular}
\end{table}

\begin{table}
\caption{Line intensities for Model set 1c}   
\label{fluxes_1c}    
\centering         
\begin{tabular}{l c c c c c}    
\hline\hline                
Line       & \multicolumn{5}{c}{Cosmic ray rate (s$^{-1}$)}  \\  
           & 5(-17) & 5(-16) & 5(-15) & 5(-14) & 5(-13) \\
\hline                     
\multicolumn{6}{c}{Line intensities (erg~s$^{-1}$~cm$^{-2}$~sr$^{-1}$)} \\
\hline
$[$CII$]$ 158~$\mu$m & 1.1(-3) & 1.1(-3) & 1.1(-3) & 1.1(-3) & 1.3(-3) \\ 
$[$CI$]$ 610~$\mu$m  & 8.3(-7) & 1.3(-6) & 2.5(-6) & 3.8(-6) & 7.7(-6) \\  
$[$CI$]$ 370~$\mu$m  & 7.1(-6) & 1.1(-5) & 2.1(-5) & 3.2(-5) & 6.3(-5) \\ 
$[$OI$]$ 63~$\mu$m   & 1.1(-1) & 1.1(-1) & 1.1(-1) & 1.3(-1) & 1.5(-1) \\ 
$[$OI$]$ 146~$\mu$m  & 3.9(-3) & 3.9(-3) & 4.1(-3) & 5.4(-3) & 6.7(-3) \\ 
$[$NII$]$ 205~$\mu$m & 7.3(-11) & 7.3(-10) & 7.5(-9) & 7.1(-8) & 4.7(-7) \\    
$[$NII$]$ 122~$\mu$m & 1.1(-9) & 1.1(-8) & 1.1(-7) &  1.1(-6) & 7.4(-6) \\
H$_3^+$ 95~$\mu$m & 1.8(-9) & 1.7(-8) & 1.8(-7) & 2.1(-6) & 9.9(-6) \\
HCN $J=1-0$ & 3.9(-5) &4.0(-5) &2.2(-5)  & 4.7(-6) & 1.8(-7) \\
HCN $J=2-1$ & 2.1(-5) &2.0(-5) &1.7(-5)  & 9.5(-6) & 2.7(-6) \\ 
HCN $J=3-2$ & 1.8(-5) &1.9(-5) &1.8(-5)  & 1.4(-5) & 4.1(-6) \\
HCN $J=4-3$ & 3.0(-5) &3.1(-5) & 3.0(-5) & 2.6(-5) & 1.2(-5) \\
HCN $J=5-4$ & 5.2(-5) &5.3(-5) & 5.3(-5) & 4.4(-5) & 2.0(-5) \\
HCN $J=6-5$ & 9.1(-5) &9.3(-5) & 9.3(-5) & 7.5(-5) & 3.5(-5) \\
HCN $J=7-6$ & 1.5(-4) &1.6(-4) &1.6(-4)  & 1.2(-4) & 5.4(-5) \\
HNC $J=  1- 0$ & 7.9(-7) & 6.0(-7) & 4.4(-7) & 1.2(-7) & 1.7(-8) \\
HNC $J=  2- 1$ & 3.3(-6) & 2.9(-6) & 2.4(-6) & 1.1(-6) & 3.0(-7) \\
HNC $J=  3- 2$ & 6.4(-6) & 5.9(-6) & 5.2(-6) & 4.4(-6) & 1.0(-6) \\
HNC $J=  4- 3$ & 1.3(-5) & 1.2(-5) & 1.1(-5) & 7.4(-6) & 2.1(-6) \\
HNC $J=  5- 4$ & 2.4(-5) & 2.2(-5) & 2.0(-5) & 1.3(-5) & 3.1(-6) \\
HNC $J=  6- 5$ & 4.1(-5) & 3.8(-5) & 3.4(-5) & 2.0(-5) & 3.4(-6) \\
HNC $J=  7- 6$ & 6.8(-5) & 6.2(-5) & 5.5(-5) & 3.0(-5) & 3.3(-6) \\
HCO$^+$ $J=  1- 0$ & 4.9(-9) & 1.0(-8) & 3.9(-8) & 1.9(-7) & 4.9(-7) \\
HCO$^+$ $J=  2- 1$ & 1.4(-7) & 2.7(-7) & 8.8(-7) & 3.1(-6) & 7.6(-6) \\
HCO$^+$ $J=  3- 2$ & 6.6(-7) & 1.2(-6) & 3.2(-6) & 8.7(-6) & 1.7(-5) \\
HCO$^+$ $J=  4- 3$ & 1.7(-6) & 2.7(-6) & 6.8(-6) & 1.7(-5) & 3.1(-5) \\
HCO$^+$ $J=  5- 4$ & 2.7(-6) & 4.2(-6) & 1.2(-5) & 3.0(-5) & 5.0(-5) \\
HCO$^+$ $J=  6- 5$ & 3.2(-6) & 5.1(-6) & 1.7(-5) & 4.7(-5) & 7.9(-5) \\
HCO$^+$ $J=  7- 6$ & 3.3(-6) & 5.2(-6) & 2.2(-5) & 7.0(-5) & 1.2(-4) \\
CO $J=  1- 0$ & 2.1(-7) & 2.3(-7) & 2.7(-7) & 2.9(-7) & 2.3(-7) \\ 
CO $J=  2- 1$ & 5.5(-6) & 5.9(-6) & 6.7(-6) & 6.9(-6) & 5.7(-6) \\ 
CO $J=  3- 2$ & 3.1(-5) & 3.3(-5) & 3.6(-5) & 3.7(-5) & 3.2(-5) \\ 
CO $J=  4- 3$ & 9.6(-5) & 1.0(-4) & 1.1(-4) & 1.1(-4) & 9.8(-5) \\ 
CO $J=  5- 4$ & 2.2(-4) & 2.2(-4) & 2.4(-4) & 2.4(-4) & 2.2(-4) \\ 
CO $J=  6- 5$ & 4.1(-4) & 4.2(-4) & 4.4(-4) & 4.5(-4) & 4.2(-4) \\ 
CO $J=  7- 6$ & 6.8(-4) & 7.0(-4) & 7.3(-4) & 7.5(-4) & 7.0(-4) \\ 
CO $J=  8- 7$ & 1.0(-3) & 1.1(-3) & 1.1(-3) & 1.2(-3) & 1.1(-3) \\ 
CO $J=  9- 8$ & 1.5(-3) & 1.5(-3) & 1.6(-3) & 1.6(-3) & 1.6(-3) \\ 
CO $J= 10- 9$ & 2.0(-3) & 2.1(-3) & 2.2(-3) & 2.2(-3) & 2.1(-3) \\ 
CO $J= 11-10$ & 2.6(-3) & 2.7(-3) & 2.9(-3) & 2.9(-3) & 2.8(-3) \\ 
CO $J= 12-11$ & 3.3(-3) & 3.4(-3) & 3.6(-3) & 3.7(-3) & 3.6(-3) \\ 
CO $J= 13-12$ & 4.0(-3) & 4.1(-3) & 4.4(-3) & 4.5(-3) & 4.5(-3) \\ 
CO $J= 14-13$ & 4.7(-3) & 4.9(-3) & 5.2(-3) & 5.4(-3) & 5.4(-3) \\ 
CO $J= 17-16$ & 6.5(-3) & 6.8(-3) & 7.3(-3) & 7.6(-3) & 8.1(-3) \\ 
CO $J= 20-19$ & 6.6(-3) & 7.0(-3) & 7.7(-3) & 8.3(-3) & 9.6(-3) \\ 
CO $J= 24-23$ & 1.8(-3) & 2.1(-3) & 2.9(-3) & 3.7(-3) & 6.3(-3) \\ 
CO $J= 30-29$ & 3.7(-5) & 4.2(-5) & 5.4(-5) & 6.9(-5) & 2.1(-4) \\ 
CO $J= 35-34$ & 3.1(-6) & 3.2(-6) & 3.5(-6) & 3.6(-6) & 7.6(-6) \\ 
$^{13}$CO $J=  1- 0$ & 5.5(-9) & 6.1(-9) & 7.2(-9) & 7.6(-9) & 5.6(-9) \\
$^{13}$CO $J=  2- 1$ & 1.7(-7) & 1.9(-8) & 2.3(-7) & 2.4(-7) & 1.7(-7) \\
$^{13}$CO $J=  3- 2$ & 1.2(-6) & 1.4(-6) & 1.6(-6) & 1.7(-6) & 1.3(-6) \\
$^{13}$CO $J=  4- 3$ & 5.0(-6) & 5.5(-6) & 6.5(-6) & 6.8(-6) & 5.0(-6) \\
$^{13}$CO $J=  5- 4$ & 1.4(-5) & 1.5(-5) & 1.8(-5) & 1.9(-5) & 1.4(-5) \\
$^{13}$CO $J=  6- 5$ & 3.2(-5) & 3.5(-5) & 4.1(-5) & 4.3(-5) & 3.2(-5) \\
$^{13}$CO $J=  7- 6$ & 6.1(-5) & 6.6(-5) & 7.8(-5) & 8.2(-5) & 6.3(-5) \\
$^{13}$CO $J=  8- 7$ & 1.0(-4) & 1.1(-4) & 1.3(-4) & 1.4(-4) & 1.1(-4) \\
$^{13}$CO $J=  9- 8$ & 1.6(-4) & 1.7(-4) & 2.0(-4) & 2.1(-4) & 1.7(-4) \\
$^{13}$CO $J= 10- 9$ & 2.2(-4) & 2.4(-4) & 2.8(-4) & 3.0(-4) & 2.4(-4) \\
$^{13}$CO $J= 11-10$ & 2.9(-4) & 3.1(-4) & 3.6(-4) & 3.9(-4) & 3.3(-4) \\
$^{13}$CO $J= 12-11$ & 3.5(-4) & 3.8(-4) & 4.4(-4) & 4.8(-4) & 4.2(-4) \\
$^{13}$CO $J= 13-12$ & 4.0(-4) & 4.3(-4) & 5.1(-4) & 5.5(-4) & 5.1(-4) \\
$^{13}$CO $J= 14-13$ & 4.2(-4) & 4.6(-4) & 5.4(-4) & 6.0(-4) & 5.9(-4) \\
\hline                           
\end{tabular}
\end{table}

\begin{table}
\caption{Line intensities for Model set 2}   
\label{fluxes_2}    
\centering         
\begin{tabular}{l c c c c c}    
\hline\hline                
Line       & \multicolumn{5}{c}{Cosmic ray rate (s$^{-1}$)}  \\  
           & 5(-17) & 5(-16) & 5(-15) & 5(-14) & 5(-13) \\
\hline                     
\multicolumn{6}{c}{Line intensities (erg~s$^{-1}$~cm$^{-2}$~sr$^{-1}$)} \\
\hline
$[$CII$]$ 158~$\mu$m & 4.4(-4) & 4.7(-4) & 5.8(-4) & 8.6(-4) & 1.2(-3) \\
$[$CI$]$ 610~$\mu$m  & 3.6(-6) & 4.6(-6) & 9.7(-6) & 6.6(-6) & 2.8(-6) \\ 
$[$CI$]$ 370~$\mu$m  & 1.0(-5) & 1.3(-5) & 3.5(-5) & 3.0(-5) & 1.5(-5) \\
$[$OI$]$ 63~$\mu$m   & 4.5(-4) & 6.4(-4) & 1.3(-3) & 2.4(-3) & 2.9(-3) \\
$[$OI$]$ 146~$\mu$m  & 5.5(-6) & 7.0(-6) & 1.1(-5) & 2.1(-5) & 3.1(-5) \\
$[$NII$]$ 205~$\mu$m & 6.6(-9) & 5.3(-8) & 3.7(-7) & 2.1(-6) & 7.6(-6) \\ 
$[$NII$]$ 122~$\mu$m & 4.2(-7) & 7.2(-7) & 5.5(-6) & 3.3(-5) & 1.1(-4) \\
H$_3^+$ 95~$\mu$m & 4.7(-10) & 7.0(-9) & 3.2(-8) & 2.0(-8) & 3.6(-9) \\
CO $J=  1- 0$ & 7.6(-8) & 1.3(-7) & 5.0(-8) & 2.8(-9) & 5.2(-12) \\
CO $J=  2- 1$ & 4.0(-7) & 6.5(-7) & 4.4(-7) & 5.3(-8) & 1.2(-10) \\
CO $J=  3- 2$ & 1.5(-6) & 2.0(-6) & 1.1(-6) & 1.4(-7) & 3.6(-10) \\
CO $J=  4- 3$ & 4.7(-6) & 5.7(-6) & 1.4(-6) & 1.4(-7) & 4.2(-10) \\
CO $J=  5- 4$ & 1.2(-5) & 1.4(-5) & 7.6(-7) & 8.9(-8) & 2.9(-10) \\
CO $J=  6- 5$ & 2.2(-5) & 2.6(-5) & 2.7(-7) & 4.1(-8) & 1.6(-10) \\
CO $J=  7- 6$ & 3.2(-5) & 4.1(-5) & 7.5(-8) & 1.6(-8) & 7.0(-11) \\
CO $J=  8- 7$ & 3.4(-5) & 5.3(-5) & 1.9(-8) & 5.3(-9) & 2.8(-11) \\
CO $J=  9- 8$ & 2.3(-5) & 5.1(-5) & 4.4(-9) & 1.6(-9) & 1.1(-11) \\
CO $J= 10- 9$ & 2.9(-6) & 2.7(-5) & 1.1(-9) & 4.9(-10) & 4.3(-12) \\
CO $J= 11-10$ & 7.2(-9) & 1.8(-6) & 3.4(-10) & 1.5(-10) & 2.0(-12) \\
CO $J= 12-11$ & 5.0(-12) & 4.0(-9) & 1.1(-10) & 5.2(-11) & 1.1(-12) \\
CO $J= 13-12$ & 3.6(-13) & 4.3(-11) & 4.0(-11) & 1.9(-11) & 6.5(-13) \\
CO $J= 14-13$ & 1.0(-13) & 2.1(-12) & 1.2(-11) & 5.8(-12) & 1.8(-13) \\
$^{13}$CO $J=  1- 0$ & 3.1(-8) & 4.6(-8) & 1.6(-9) & 5.1(-11) & 9.1(-14) \\
$^{13}$CO $J=  2- 1$ & 1.6(-7) & 2.6(-7) & 2.5(-8) & 1.2(-9) & 2.4(-12) \\
$^{13}$CO $J=  3- 2$ & 3.3(-7) & 6.4(-7) & 5.3(-8) & 3.9(-9) & 9.5(-12) \\
$^{13}$CO $J=  4- 3$ & 2.0(-7) & 7.9(-7) & 4.6(-8) & 5.0(-9) & 1.4(-11) \\
$^{13}$CO $J=  5- 4$ & 2.0(-8) & 2.5(-7) & 2.6(-8) & 3.9(-9) & 1.2(-11) \\
$^{13}$CO $J=  6- 5$ & 1.9(-9) & 4.4(-8) & 1.1(-8) & 2.3(-9) & 8.2(-12) \\
$^{13}$CO $J=  7- 6$ & 1.2(-10) & 7.5(-9) & 4.4(-9) & 1.2(-9) & 4.7(-12) \\
$^{13}$CO $J=  8- 7$ & 9.8(-12) & 1.1(-9) & 1.5(-9) & 5.2(-10) & 2.4(-12) \\
\hline                           
\end{tabular}
\end{table}

\section{Key Diagnostics}\label{key_diagn}

\begin{figure}
  \centering \includegraphics[width=8cm,clip]{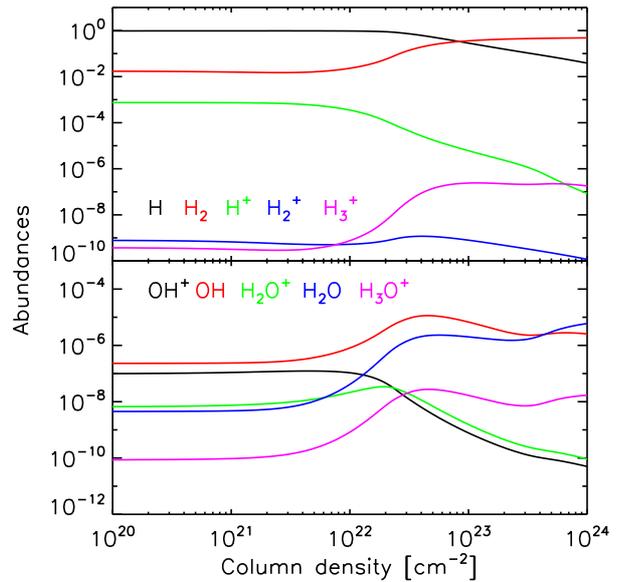}
  \caption{Abundances of hydrogen and water chemistry for an XDR model
    with density $n=10^{5.5}$~cm$^{-3}$ and
    $F_x=28.5$~erg~cm$^{-2}$~s$^{-1}$.}
  \label{XDR_model}
\end{figure}

The aim of this paper is to pinpoint potential line diagnostics that
can be used to trace an enhanced cosmic ray radiation field and
mechanical heating in the presence of starformation such as the
central regions of Arp~220 and NGC~253. One of the remarkable results
is that many commonly observed lines, such as, the [OI]~63~$\mu$m,
[CI]~609~$\mu$m, the [CII]~158~$\mu$m and the low-$J$ CO lines, show
very moderate responses to high cosmic ray rates and mechanical
feedback, when the medium is already exposed to a large amount of
UV. This is a very useful result, since these lines can serve as a
reference point to other lines fluxes, that are impacted.

\subsection{Tracers of Mechanical Heating}

Mechanical heating is only increasing the temperature of the gas,
leaving the ionization fraction unaffected. Tracing this heating
mechanism with molecules that contain activation barriers in their
formation route is thus very useful. Lines of, e.g., HCN and H$_2$O
are enhanced with respect to the aforementioned fine-structure lines,
especially for the highest considered mechanical heating rate. It was
already pointed out by \citet{Loenen2008} that the HCN/HNC line ratios
increase when mechanical heating is important, since at high
temperatures the HNC will be driven in to HCN (see also Section
\ref{Chemistry}). Moreover, the HCN/CO line ratios also increase when
mechanical heating effects are important, but the interpretation of an
observed ratio is not straightforward, because the lines trace
slightly different regions.

\subsection{Tracers of Enhanced Cosmic Ray Rates}

The [NII] fine-structure lines show a very strong response to enhanced
cosmic ray rates, independent of mechanical heating effects. The main
concern with these lines, however, is that one has to disentangle the
contributions arising from these CR exposed clouds and HII regions,
that also contribute a significant amount of [NII] emission in
starforming environments.

Most promising would be the study of lines from species important in
the water chemistry, especially when the ionized water related species
are considered. The OH to H$_2$O column density ratio varies over two
to three orders of magnitude for the cosmic ray rates considered, from
$\sim 5\cdot 10^{-3}$ ($\zeta=5.0\cdot 10^{-17}$~s$^{-1}$) to $\sim
1-3$ ($\zeta=5.0\cdot 10^{-13}$~s$^{-1}$) in the high density case,
and $\sim 0.2$ to $\sim 400$ in the low density case. So generally the
OH lines are expected to be stronger in comparison to the water lines
for higher cosmic ray rates. Fig. \ref{XDR_model} shows an XDR model
with density $n=10^{5.5}$~cm$^{-2}$, and radiation field
$F_X=28.5$~erg~cm$^{-2}$~s$^{-1}$ (a radiation field at $\sim$170~pc
distance from a typical AGN with $L_X=10^{44}$~erg/s). From this we
derive that the ratio varies between $\sim 8$ (integrated up to
$N_{\rm H}=3\cdot 10^{22}$~cm$^{-2}$) and $\sim 1$ ($N_{\rm
  H}=10^{24}$~cm$^{-2}$). So for very high CR rates, the ratios become
comparable to a typical X-ray illuminated cloud.

\citet{VdTak2008} already pointed out that the H$_3$O$^+$/H$_2$O ratio
is affected by either cosmic rays or X-rays. The high density models
show variations over 1 to 2 orders of magnitude in the high density
case. Unfortunately, the interpretation is not entirely unambiguous,
since the H$_3$O$^+$ abundance decreases for the highest CR
rates. Also, clouds illuminated by high CR rates have similar ratios
as those exposed to X-rays, namely $(1-6) \cdot 10^{-3}$
($\zeta=5\cdot 10^{-14}$~s$^{-1}$) and $(5-10)\cdot 10^{-3}$
($\zeta=5\cdot 10^{-13}$~s$^{-1}$) for models set 1, and between
$15-4\cdot 10^{-3}$ (depending on the size of the cloud) for the XDR
model shown in Fig. \ref{XDR_model}.

The current study reveals that additional observations of OH$^+$ and
H$_2$O$^+$ might help out in distinguishing very high CR rates from a
typical XDR. The OH$^+$ and H$_2$O$^+$ are significantly enhanced for
CR rates that are $10^{3}-10^{4}$ times larger than the accepted Milky
Way value, but not as much as in the XDR. OH$^+$/OH and
H$_2$O$^{+}$/H$_2$O column density ratios can be as high as
$(3-5)\cdot 10^{-4}$ to $(2-14)\cdot 10^{-4}$, respectively,
compared to $(250 - 7.5) \cdot 10^{-4}$ and $(64-4)\cdot 10^{-3}$
($N_{\rm H}=3\cdot 10^{22}-10^{24}$~cm$^{-2}$). XDRs and regions with
enhanced CR rates can be distinguished when all species are
considered.

In this paper, no attempt is made to model lines from these
species. As already pointed out in, e.g., \citet{Gonzalez2008}, the
radiation transfer of OH and H$_2$O involves the consideration of both
collisions and radiation pumping, which is beyond the scope of this
paper. Also, OH$^+$ and H$_2$O$^+$ lines are not easily modeled. These
species have formation and destruction times scales that are similar
those for excitation. Besides that, collisional cross sections are
unavailable at the moment, and quantum mechanical calculations are
needed to obtain them.

\section{Outlook}

OH$^+$ and H$_2$O$^+$ have already been detected in a number of
sources:

\begin{enumerate}

\item[-] \citet{VdWerf2010} show OH$^+$ and H$_2$O$^+$ emission lines
  in the ULIRG Mrk231, which are about 30 percent of strength of the
  CO(5-6) to CO(13-12) lines. The paper show that these lines are the
  result from AGN activity in Mrk231.

\item[-] In the Milky Way, OH$^+$ has been detected in absorption
  toward Sagittarius B2 using the Atacama Pathfinder Experiment (APEX)
  by \citet{Wyrowski2010}, finding a column density of $2.4\cdot
  10^{15}$~cm$^{-2}$. \citet{Ossenkopf2010} found H$_2$O$^+$ in
  absorption toward, NGC~6334, DR21, again Sgr B2, obtaining column
  densities of $7.2\cdot 10^{12}$~cm$^{-2}$, $2.3\cdot
  10^{13}$~cm$^{-2}$, and $1.1\cdot 10^{15}$~cm$^{-2}$,
  respectively. These galactic observations relate these species to
  the diffuse ISM.

\end{enumerate}

\noindent
Extensive modeling of the excitation of water related species is very
timely, since more and more observations will be available in a very
short timefrace from observations with the recently lauched {\it
  Herschel} Space Telescope.

\begin{acknowledgements}
We thank Padelis Papadopoulos for useful discussions. We thank the
anonymous referee for a careful reading of the manuscript and
constructive comments.
\end{acknowledgements}

\bibliographystyle{aa}

\end{document}